\documentclass[aps,prfluids,longbibliography,floatfix,10pt]{revtex4-2}
\usepackage{CJK}
\usepackage{siunitx}
\usepackage{graphicx,ifpdf}% Include figure files
\usepackage{amsmath,amssymb}
\usepackage{cases,booktabs}
\graphicspath{ {./images/} }
\usepackage{ifpdf}\ifpdf % si pdftex
    \usepackage[pdftitle={Physics and Modeling of Liquid Films in Pulsating Heat Pipes},linktocpage,pdftex,bookmarksopen,bookmarksopenlevel=1,hyperfigures]{hyperref}
\else
    
    \usepackage{hyperref}
\fi
\usepackage{lineno}

\begin{document}
\begin{CJK*}{UTF8}{gbsn}
\title{Physics and Modeling of Liquid Films in Pulsating Heat Pipes}
\author{Xiaolong Zhang (张晓龙)}
\author{Vadim S. Nikolayev}\email[Corresponding author; e-mail: ]{vadim.nikolayev@cea.fr}
\affiliation{Service de Physique de l'Etat Condens\'e, CEA, CNRS, Universit\'e Paris-Saclay, CEA Paris-Saclay, 91191 Gif-sur-Yvette Cedex, France}
\date{\today}

\begin{abstract}
The present study reports a novel physical model for simulating Pulsating Heat Pipes (PHP). Their high heat performance is due to the phase change over thin liquid films. The simulation of physically correct film behavior is thus crucial. The model adopts the one-dimensional approach, which is  computationally efficient yet still capable of capturing major physical phenomena. The model assumes a spatially uniform film thickness, whereas both the film thickness and length can vary over time; therefore, we call it the oscillating film thickness model. It is based on the physical analysis of liquid film deposition by the receding menisci of Taylor bubbles and of contact line dynamics. Three key phenomena are addressed: (i) film deposition, (ii) contact line receding due to dewetting acceleration by evaporation, and (iii) mass exchange over films and contact lines. The  model is evaluated by simulating the simplest, single-branch PHP, for which detailed experimental data are available. A quantitative agreement is reached. As the model includes the wetting properties, their impact on oscillations is analyzed; a qualitative agreement with the experiment is demonstrated.
\end{abstract}

%\pacs{47.55.np, 68.03.Fg, 47.55.nb, 68.08.Bc}
\maketitle
\end{CJK*}
%\linenumbers

\section{Introduction}
\begin{figure}[ht]
  \centering
  \includegraphics[height=4.5cm,clip]{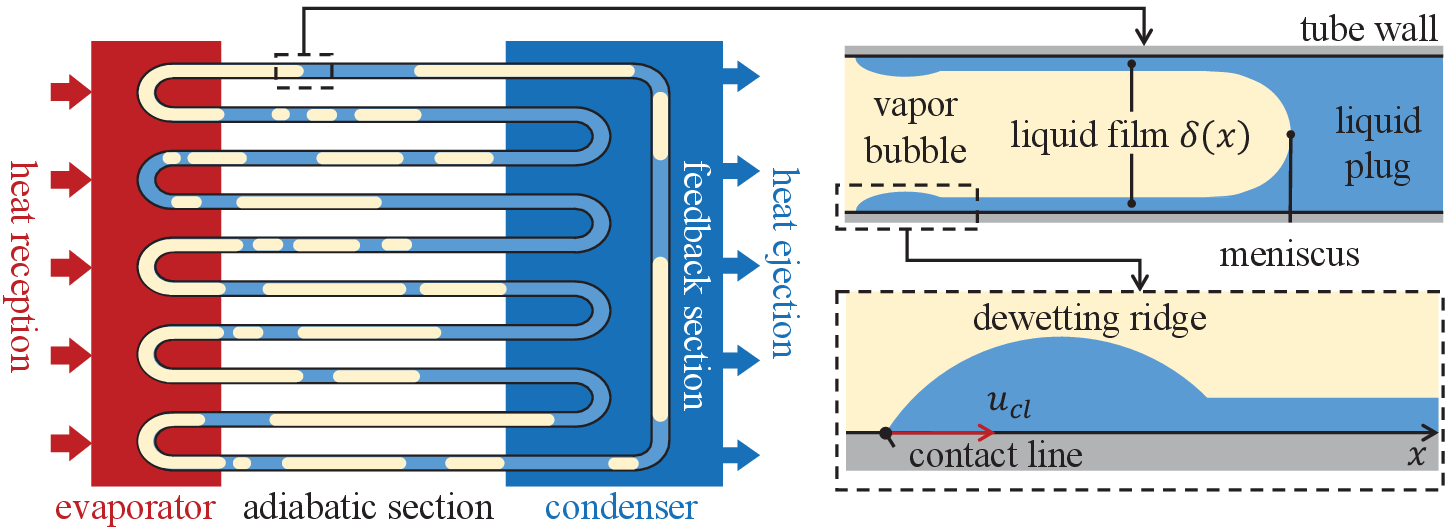}
  \caption{Schematic of a closed-loop 10-branch PHP with evaporator, condenser and adiabatic section; the inside of a branch is demonstrated: liquid plug, meniscus, liquid film, contact line, and vapor bubble.}\label{fig:closedLoopPHP}
\end{figure}

Invented in the 1990s , the Pulsating Heat Pipe (PHP) has proven to be a promising alternative to conventional heat pipes \cite{ATE21}. Numerical simulations and experimental measurements have confirmed the high heat transfer capacity of PHP, making it an attractive choice for various applications, such as electronics and renewable energy. The PHP structure is simple: it is a capillary tube that meanders between a heater and a cooler, with the heated sections referred to as evaporators and the cooled sections, as condensers. Adiabatic sections separating the evaporators and condensers can also be present, see Fig.~\ref{fig:closedLoopPHP}. The PHP is charged with a pure two-phase fluid, which forms a sequence of vapor bubbles and liquid plugs within the closed capillary tube. The most important parameter of PHP is the number of branches; the branch is a tube segment connecting the evaporator and the condenser. For instance, Fig.~\ref{fig:closedLoopPHP} illustrates a PHP of ten branches.

Given a proper temperature difference between evaporators and condensers, the cyclic phase change (liquid evaporation and condensation) can engender self-sustained oscillation of liquid plugs. This continuous flow of liquid within the PHP leads to efficient heat transfer.  The numerical simulations \cite{JHT11,MST19} and in-situ experimental measurements \cite{Jo19,Kamijima20,Francom21} have revealed that the heat transfer mechanism in a PHP is predominantly through the latent heat transfer via thin liquid films (typical thickness $\delta \sim \SI{100}{\micro m}$) deposited by the receding liquid menisci. The \emph{in situ} experimental studies \cite{PRF16,LauraATE17} evidence the liquid ridges developed due to contact lines receding towards the liquid side as schematized in Fig.~\ref{fig:closedLoopPHP}.

Despite the outstanding performance and design simplicity of PHPs, the underlying mechanisms are not yet fully understood, and the available models are of insufficient quality, which hampers the industrial application of this technology. Some researchers proposed using empirical correlations \cite{Khandekar03b,Qu13}, which is a conventional approach in heat pipe design. Recent attempts to use artificial neural networks  \cite{Jokar16, Jalilian16} to predict PHP performance have been reliable only within the range of experimental parameters for which the models were established. To obtain a predictive PHP design tool, numerical simulations based on physical models and understanding the effects inside capillary tubes remain the most promising approach. These simulations are particularly valuable in cases where the PHP operates under conditions that are difficult or impossible to replicate experimentally. Currently, multi-dimensional simulations are not yet reliable due to substantial difficulties in describing the micrometric films and meter-scale hydrodynamics within the same simulation \cite{ATE21}. Several one-dimensional physical models have been proposed, and one of the primary differences among these models lies in their treatment of  the liquid film, which plays the most crucial role in dictating the heat transfer characteristics of the device.

\begin{figure}
  \centering
  \includegraphics[width=14cm,clip]{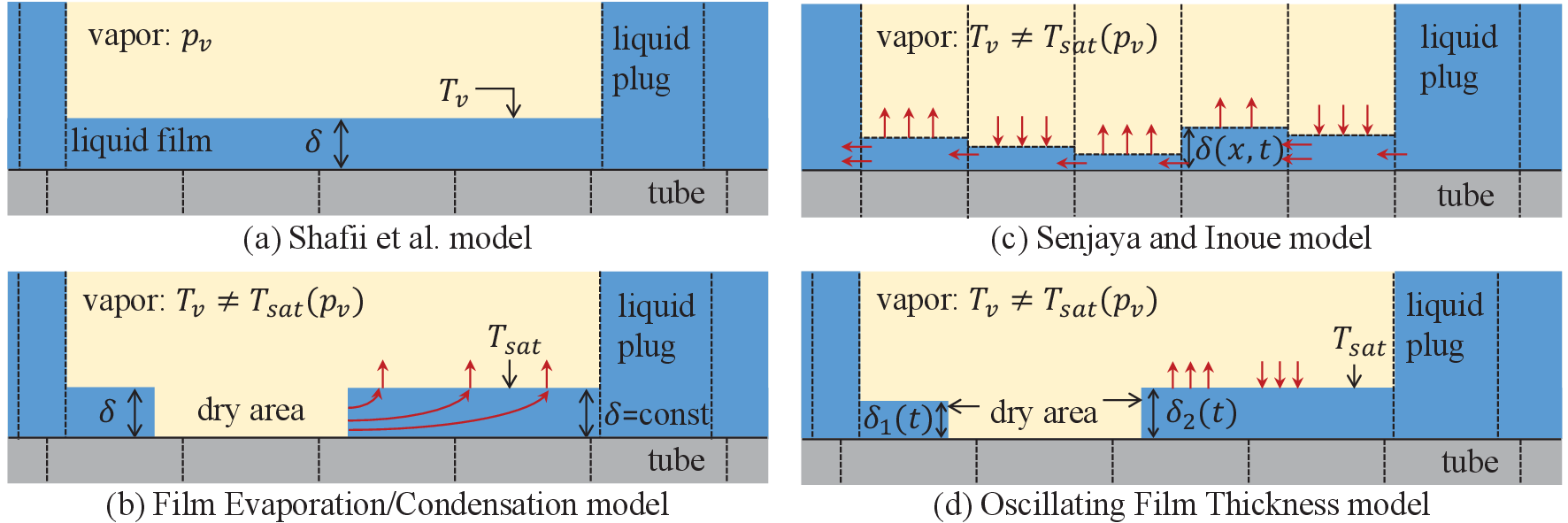}
  \caption{Major features and differences of existing one-dimensional PHP models for plug/slug flow in PHP. The tube radial cross-section is schematized.}\label{fig:PHPmodels}
\end{figure}

In the pioneering work of \citet{shafii1} and subsequent models based on it (\cite{Holley05,Mameli12} to cite a few), it was assumed that vapor bubbles are surrounded by a flat liquid film of uniform and constant thickness (see Fig.~\ref{fig:PHPmodels}a). On the one hand, the vapor phase was considered to remain superheated with respect to the saturation temperature $T_{sat}$ for the current vapor pressure $p_{v}$ and was described using the ideal gas law. The heat transfer through the films is defined by the difference between the tube wall temperature $T_{w}$ and the vapor temperature $T_{v}$. However, the heat transfer provided by thin liquid films should be proportional to $(T_{w}-T_{sat})/\delta$  \cite{PF10}, indicating that $T_{sat}=T_{v}$ is assumed implicitly in this model, which is self-contradictory. Moreover, since in this model the heat exchange is $\sim(T_{w}-T_{v})$, i.e. the convective heat exchange between the vapor and the dry tube wall, the liquid films can be considered as neglected, which leaves to the vapor a key role in dynamics. For this reason, this model is referred to as the ``Superheated vapor'' model \cite{IJHMT10}. Experimental work \citet{Gully14,Rao13} has confirmed that, in a functioning PHP, the vapor stays indeed superheated so it can be described as ideal gas. Simplicity is the major advantage of the superheated vapor model, but it cannot describe the film drying, which is a major phenomenon providing the PHP functioning limit. In reality, liquid films only partially cover the tube wall during the PHP functioning. They vary with time, which is an essential factor of the PHP dynamics. An approach \cite{Tessier-Poirier19} similar to the superheated vapor model has been recently explored aiming to explain the origin of the self-sustained oscillation in PHP. However, the superheated vapor model can produce only small-amplitude oscillations, in many cases much weaker than those observed experimentally.

To overcome this limitation, \citet{IJHMT10} introduced the time change of liquid films in what they called ``Film Evaporation/Condensation'' (FEC) model to describe the simplest, single-branch PHP (see Fig.\ref{fig:SBPHP} below). The model defines the mass exchange in terms of the temperature difference between the wall temperature $T_{w}$ and the film interfacial temperature $T_{sat}$. The vapor bulk temperature $T_{v}$ differs from $T_{sat}$ (more precisely, larger than $T_{sat}$), meaning that there is a thin temperature boundary layer in vapor near the film interface (which is justified by low thermal diffusivity of vapor). The FEC model also included partial film drying, commonly observed in experiments. The model assumed a constant imposed film thickness $\delta$, while allowing for variations in the film length in response to overall film evaporation to conserve its mass (see Fig.~\ref{fig:PHPmodels}b and sec.~\ref{ResFEC} below). However, this assumption implicitly implies a flow along the film (Fig.~\ref{fig:PHPmodels}b), which is not representative of reality. The flow along thin liquid films is extremely slow. A deformed film of typical thickness 10-\SI{100}{\micro m} in PHPs \cite{ATE21} does not have time to relax between the back-and-forth motions of liquid plugs \cite{JFM21}. Once deposited, films can thus be regarded as ``frozen''. Therefore, a physically justified model should avoid film flow along the film length.

\citet{JHT11} generalized the FEC model to the multi-branch PHPs and implemented the model in the software CASCO (the French abbreviation for Advanced PHP simulation code: Code Avanc\'e de Simulation de Caloduc Oscillant) in subsequent works \cite{IarATE17,MST19,IJHMT22}. The numerical results reproduced self-sustained, sometimes chaotic large-amplitude oscillations of liquid plugs observed in experiments; several oscillation regimes were identified. However, one of the objectionable features of the FEC model remains the necessity of choosing an \emph{a priori} fixed value for $\delta$.

For multi-branch PHP simulations, \citet{dEntremont15,Nemati18} introduced the ``wedge film'' model where the film thickness $\delta(x,t)$ was variable. Its variation is defined by the local evaporation rate proportional to $[T_{w}(x,t)-T_{sat}]/\delta(x,t)$. As this rate diverges at $\delta\to 0$, they introduced a minimum thickness $\delta_{min}$, below which the evaporation halts. The PHP evolution is expected to be strongly affected by choice of $\delta_{min}$ as it limits the mass exchange in the system.

In addition to the film thinning discussed above, \citet{Senjaya14} explicitly introduced axial flows between control volumes in films (Fig.~\ref{fig:PHPmodels}c) in their 1D model. The film flow is induced rather by the film thickness gradient than the film curvature gradient that drives the film flow in reality. The film flow is coupled to the heterogeneous pressure calculation in the vapor phase. Consequently, the required computer resources (time, memory) are much larger than for all the above models. The computational grid of the mesh size $\sim\SI{1}{mm}$ is common for the liquid film, liquid plug and vapor domains (Fig.~\ref{fig:PHPmodels}c). However, such a mesh size is far from being sufficiently small to describe the curved menisci, which is essential to describe the physically realistic film dynamics \cite{JFM21}. The \citeauthor{Senjaya14} model has been compared to the FEC and Superheated vapor models by \citet{Bae17}. Both \citeauthor{Senjaya14} and FEC models produced oscillations and thermal behavior similar to those observed experimentally (provided an appropriate choice of $\delta$ for the FEC model). However, the Superheated vapor model failed to produce oscillating behavior.

The wettability has a significant impact on the PHP performance \cite{Hao14}. However, to the best knowledge of the authors, none of the existing models accounts for it.

Physical analysis of the PHP was attempted by many authors but is still incomplete. In some models \cite{ATE21}, PHP is described with a mass-spring model where the liquid plugs are masses and the superheated vapor plays the spring role. This approach is suitable for the micro-PHP \cite{Yoon19}, where viscous effects are so strong that the oscillations are of small amplitude, and for the single-branch PHP where there is only one plug, i.e., a unique eigenfrequency \cite{IJHMT10}. For this latter case, the frequency scaling is discussed below (\autoref{secvap}).

For larger amplitudes, the vapor amount strongly varies during oscillations, which causes strong nonlinear effects and dynamical chaos.

The instability that causes the oscillations in PHP was understood \cite{ATE21} for the single-branch PHP. The origin of oscillations was clearly identified as a phase change appearing when the meniscus deviates from the neutral position. This leads to the vapor pressure change that pushes the meniscus backwards tending to restore the neutral position. The liquid plug inertia creates a phase shift between the pressure force and meniscus position, which is well known to be necessary to cause self-sustained oscillations. Evidently, the amplitude of this deviation grows in time when the energy brought into the system is larger than dissipation. The instability analysis was performed to define the threshold of self-sustained oscillations in terms of the heating power applied to the evaporator. It was shown within the FEC model \cite{IJHMT16,ATE21} that this threshold is a sum of three contributions that represent the viscous losses, the thermal losses and a term proportional to the slope of saturation curve (that represents the vapor mass variation that hinders its elastic response). Similar results were obtained later within the superheated vapor model \cite{Tessier-Poirier19}.

However, the main challenge of understanding PHP is its stable functioning mode (the most interesting for applications) controlled by nonlinearities. The most important of them is probably caused by the very existence of liquid films \cite{IJHMT16}. When a liquid plug advances over the dry wall, heat and mass exchange occurs only from a small portion (adjacent to its contact line) of the front plug meniscus. In contrast, when the plug starts to recede, the deposited liquid film provides much more efficient exchange. Such a dissymmetry between advancing and receding leads to a nonlinearity in the model even for small-amplitude oscillations. This is why a precise and physically sound description of the liquid film is crucial to understand the PHP dynamics. This work aims to develop a PHP model that balances computational efficiency and the liquid film physical description that accounts for the recent findings on both the thin film formation \cite{Snoeijer06a,Gao19} and their behavior \cite{JFM21,JFM22}. The proposed model (that we call ``Oscillating Film Thickness''  model abbreviated to OFT to distinguish from previous approaches) is validated using experimental data on the single-branch PHP, for which detailed observations of fluid behavior, such as contact line and meniscus dynamics and vapor pressure variation in a specific bubble, are available.

The paper is structured as follows. The next section recapitulates the cornerstone physical principles for liquid films in capillary tubes with an application to PHP. Based on it, sec.~\ref{modelsec} details the development of the OFT model, followed by the presentation of the asymptotic analysis in sec.~\ref{asympRes}. Finally, the numerical results are discussed, and a comparison with the experimental data is made in sec.~\ref{ResSec}.

\section{Physics of liquid films in PHP}

Several key phenomena are relevant to the liquid film dynamics in PHP: (i) the film deposition by receding liquid plugs, the velocity of which controls the initial (``deposited'') thickness, (ii) the film evaporation and condensation affecting the film thickness, and (iii) the contact line receding that shortens the film. In this section, we focus on the spatial variation of the film thickness, denoted as $\delta=\delta(x)$, which tends to zero at the contact line as shown in Fig.~\ref{fig:closedLoopPHP}.

\subsection{Thickness of deposited liquid films}

The initial thickness $\delta_{dep}$ of a liquid film deposited by a liquid plug receding at speed $u_{m}$ can usually be approximated by the semi-empirical formula \cite{Aussillous}
\begin{equation}\label{eq:delta_dep}
  \delta_{dep}=\frac{0.67d Ca_{m}^{2/3}}{1+3.35 Ca_{m}^{2/3}},
\end{equation}
where $d$ is the inner diameter of tube, and $Ca_{m}=\mu u_{m}/\sigma$ is the meniscus capillary number. Here, $\mu$ and $\sigma$ are the liquid shear viscosity and the surface tension of vapor-liquid interface, respectively. As demonstrated previously \cite{JFM21}, this formula conforms to experiments on meniscus oscillation when Reynolds number $\lesssim 500$.

\subsection{Multiscale approach and contact angles} \label{sec:multi}

\begin{figure}[htb]
\centering
\includegraphics[width=9cm,clip]{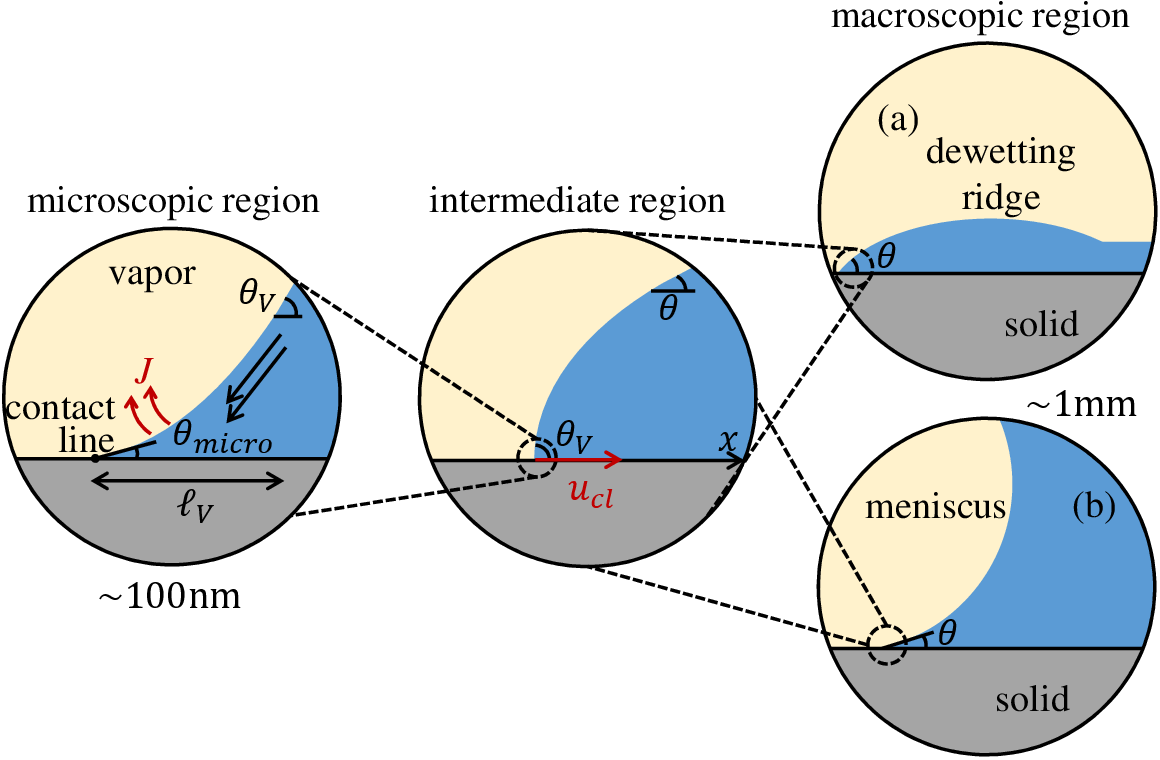}
\caption{Length scale hierarchy in the case of a moving contact line for a dewetting ridge (a) and a meniscus (b).}
\label{fig:MultiSacles}
\end{figure}

The analysis of liquid films with moving contact lines under evaporation/condensation conditions is more complex than in statics. However, due to substantial scale separation, the physical effects that govern the dynamics on one scale can be neglected at another scale \citep{SWEP22}. Generally, three regions, as illustrated in Fig.~\ref{fig:MultiSacles}, can be identified based on the distance from the contact lines \citep{PRE13movingCL}. These regions will be considered henceforth.

The microregion is the smallest region characterized by a typical scale of $\ell_{V} \sim\SI{100}{nm}$, also known as the Voinov length (see Fig.~\ref{fig:MultiSacles}a for a schematic). In the partial wetting case, the microscopic contact angle $\theta_{micro}$, which is the interfacial slope at the triple-phase contact line at a scale of $\sim\SI{1}{nm}$, is used to characterize the wetting properties of the tube wall. In the microregion, the interface curvature is predominantly controlled by the fluid flow associated with the interfacial phase change. The scaling analysis shows that the contact line motion term is negligible \cite{PRE13movingCL}. As evaporation induces a strong liquid flow towards the contact line, there are pressure gradients that lead to a highly curved interface in this region. Therefore, the Voinov angle $\theta_{V}$, to which the interface slope saturates at the right-hand boundary of the microregion (far from the contact line), is greater than $\theta_{micro}$. The value of $\theta_{V}$ depends on the microscopic contact angle $\theta_{micro}$ and the superheating $\Delta T_{cl}=T_{w}(x_{cl})-T_{sat}$ at the contact line location $x_{cl}$ \citep{EuLet12}.

Liquid flow induced by contact line motion controls the interface slope in the intermediate region, typically within the range of \SI{100}{nm} to \SI{100}{\micro\meter} from the contact line. The evaporation-induced flow in the microregion is negligible \cite{PRE13movingCL}; the balance of surface tension and viscosity governs the interfacial shape. Many researchers have extensively studied this problem. The apparent contact angle $\theta$ (i.e. the experimentally observable value) is a function of contact line velocity $u_{cl}$ \cite{Voinov, cox86},
\begin{equation}\label{eq:Cox-Voinov}
  \theta^3 = \theta_{V}^3-9Ca\ln \frac{L}{\ell_{V}},
\end{equation}
where the dimensionless contact line speed (the capillary number) is $Ca=\mu u_{cl}/\sigma$, positive for receding and negative for advancing over dry surfaces. Two parameters, $\ell_{V}$ and $\theta_{V}$, which appear as integration constants, are obtained from the asymptotic matching to the microregion solution. The characteristic length scale $L$ of the macroscopic region is dependent on the macroscopic interface geometry. In the case of Fig.~\ref{fig:MultiSacles}a, $L$ is proportional to the width of the dewetting ridge \cite{Snoeijer10}. For the ``bare'' meniscus case (Fig.~\ref{fig:MultiSacles}b) that can occur under some conditions discussed below, $L$ is proportional to the meniscus radius of curvature.

\subsection{Mass exchange in liquid films}\label{sec:Jcl}

\begin{figure}[ht]
  \centering
  \includegraphics[width=4.5cm,clip]{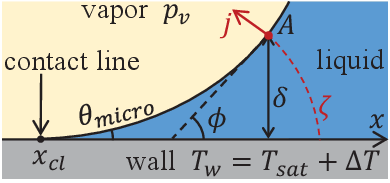}
  \caption{Sketch of a curved film with a contact line.}
  \label{fig:wedge}
\end{figure}

Fig.~\ref{fig:wedge} sketches the 2D cross-section of a liquid film in the tube. The film is bounded from above by the pure vapor of the same fluid, which has a homogeneous pressure $p_{v}$. The temperature of the inner tube wall $T_{w}$ is above the saturation temperature $T_{sat}$, which corresponds to $p_{v}$. The generalized lubrication theory \cite{JFM22} is used to describe the hydrodynamics and heat exchange in liquid films with large interface slopes (over $30^\circ$).

At an interfacial point A, it is possible to define a corresponding straight wedge. This wedge is formed by the tangent to the interface at A and the solid surface, with the wedge opening angle equal to the local slope of the interface, denoted by $\phi$. An intercepted arc is defined with a central angle of $\phi$, whose length from point A to the tube wall is
\begin{equation}\label{zeta}
\zeta=\delta \phi/\sin \phi,
\end{equation}
where $\delta$ is the local liquid thickness.

Because liquid films are thin, heat conduction is the primary mechanism for heat exchange in the radial direction. The conduction can be assumed stationary due to the low thermal inertia of thin films. The temperature variation due to conduction in the film is linear along the intercepted arc, like in a straight wedge \cite{Anderson1994}. The interfacial heat flux thus reads
\begin{equation}\label{eq:qli}
  q_{l}^{i}=k_{l}\frac{(T_{w}-T^{i})}{\zeta},
\end{equation}
where $T^{i}\approx T_{sat}$ is the interfacial temperature; $k_{l}$ is the liquid thermal conductivity. Energy balance at the interface of the temperature $T^{i}$ can thus be written as
\begin{equation}\label{eq:j}
{\cal L}j =q_{l}^{i},
\end{equation}
where $j$ is the local interfacial mass flux assumed positive at evaporation; $\cal {L}$ is the latent heat. Eq.~\eqref{eq:j} neglects the vapor-side heat flux because of the low vapor heat conduction.

As the thickness $\delta$ tends to zero at the contact line, $j\propto\zeta^{-1}$ becomes infinite. Relaxing this singularity requires accounting for several nanoscale effects that significantly affect the liquid hydrodynamics in the microregion. They are hydrodynamic slip, the Kelvin effect, interfacial thermal resistance, vapor recoil, and the Marangoni effect \cite{JFM22}. Of these, the interfacial thermal resistance $R^{i}$ \cite{SWEP22} was found to have a dominant impact on the heat flux. It is important when $j$ is so high that the corresponding velocity of vapor molecules is comparable to the thermal velocity. Because of $R^{i}$, $T^{i}$ deviates from  $T_{sat}$,
\begin{equation}\label{eq:Tint}
T^{i} = T_{sat} +  R^{i} {\cal L}j,
\end{equation}
where
\begin{equation}\label{eq:Ri}
    R^{i} = \frac{T_{sat} \sqrt{2\pi R_{v} T_{sat}} (\rho_{l}-\rho_{v}) }{2{\cal L}^2 \rho_{v}\rho_{l}},
\end{equation}
where $R_{v}$ is the specific gas constant; $\rho_{v}$ and $\rho_{l}$ are the densities of vapor and liquid, respectively. Combining Eqs. (\ref{eq:qli}--\ref{eq:Tint}) yields
\begin{equation}\label{eq:jR}
j = \frac{k_{l}\Delta T}{{\cal L}(R^{i} k_{l} + \zeta)},
\end{equation}
where $\Delta T= T_w-T_{sat}$ is the tube superheating, and the length $R^{i} k_{l}$ can be deemed as an additional layer of liquid, which is typically smaller than \SI{10}{nm}. Evidently, the interfacial resistance plays a significant role for  thin films, in particular in the contact line vicinity.

\begin{figure}[ht]
  \centering
  \includegraphics[width=5.5cm,clip]{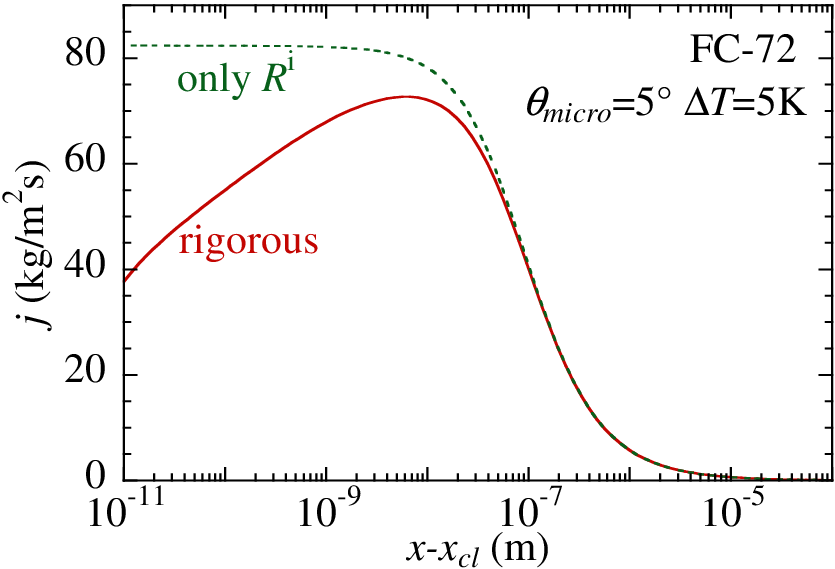}
  \caption{Typical spatial variation of mass flux along the vapor-liquid interface computed for FC-72, $\theta_{micro}=5^\circ$ and $\Delta T=\SI{5}{K}$. The other parameters are given in \autoref{tab:expsum} below.}
  \label{fig:jcl}
\end{figure}

The green dashed curve in Fig.~\ref{fig:jcl} shows $j$ given by Eq.~\eqref{eq:jR} as a function of the distance from the contact line. It can be compared to the rigorous solution \cite{JFM22} accounting for all the nanoscale effects mentioned above shown as the red curve in Fig.~\ref{fig:jcl}. The rigorous $j$ value at the contact line is proportional to its velocity and is zero when the speed is zero \cite{EuLet12}, while Eq.~\eqref{eq:jR} leads to a larger value. However, both curves coincide above \SI{100}{nm}, i.e., beyond the microregion. One is usually interested only in the total mass flux $J$, which is $j$ integrated over the interfacial area. The distribution \eqref{eq:jR} is precise enough for this purpose because the contribution of small scales is negligible.  %For instance, integrating $j(x)$ from the contact line to $(x-x_{cl})=\SI{0.1}{mm}$ gives the total mass exchange of $\approx \SI{4.74e-5}{kg/(m.s)}$ for the rigorous solution, and \SI{4.77e-5}{kg/(m.s)} for the distribution \eqref{eq:jR}. These two integrals are very close.
Therefore, instead of using the rigorous solution, our PHP model employs Eq.~\eqref{eq:jR} to evaluate $J$ in sec. \ref{MassExMen} below.

\subsection{Two interface geometries in PHP}\label{secrec}
\begin{figure}[ht]
  \centering
  \includegraphics[width=5cm,clip]{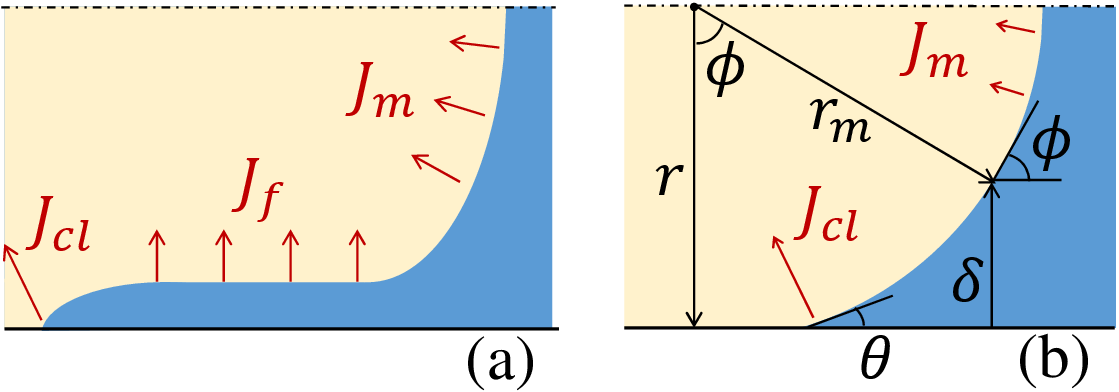}
  \caption{Schemes of two possible geometries of the vapor-liquid interface (radial cross-section) and the mass exchange over it: (a) meniscus with the deposited film and (b) in the absence of the deposited film.}
  \label{fig:dotm_menisci}
\end{figure}

Two different interface geometries schematized in Figs.~\ref{fig:dotm_menisci} are possible.

\subsubsection{Meniscus with deposited liquid film}\label{MenDep}

\begin{figure}[ht]
  \centering
  \includegraphics[width=6cm,clip]{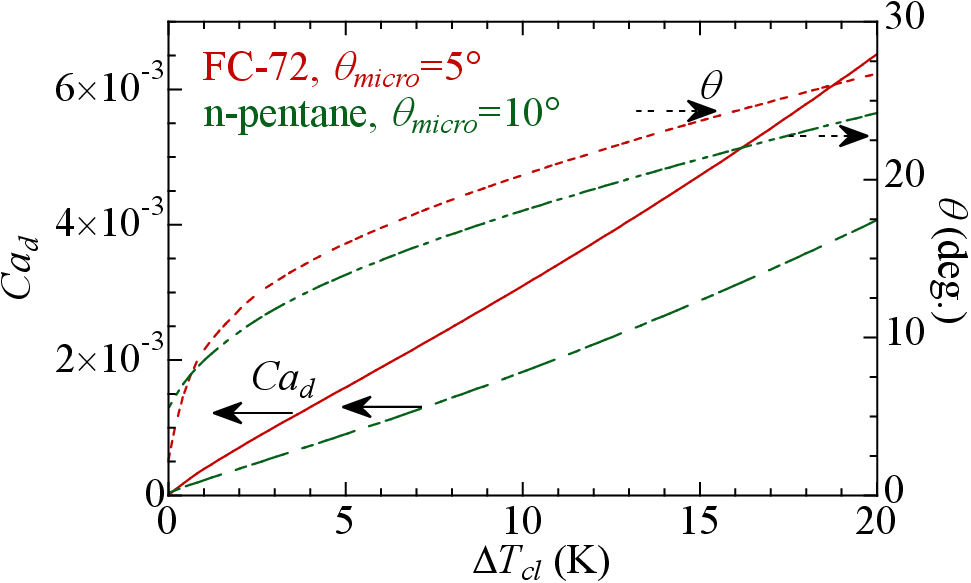}
  \caption{Dewetting capillary number $Ca_{d}=\mu u_{d}/\sigma$ and apparent contact angle for FC-72 with $\theta_{micro}=5^\circ$ (red lines), and  n-pentane with $\theta_{micro}=10^\circ$ (green lines) for the conditions of \autoref{tab:expsum} below; the curves are calculated as detailed in \cite{EPL23}.}\label{fig:FC72SBPHP}
\end{figure}

As shown in Fig.~\ref{fig:dotm_menisci}a, when a liquid film is deposited, a contact line exists at its edge. Capillary action causes the contact line to spontaneously recede, leading to the formation of a dewetting ridge (Fig.~\ref{fig:MultiSacles}a) near the contact line due to the strong viscous shear that prevents the liquid from flowing into the film \cite{SWEP22}. The contact line velocity $u_{cl}$ is given by the dewetting speed $u_{d}$, which is controlled by $\theta_{V}$. In the absence of mass exchange, $\theta_{V}=\theta_{micro}$ \cite{PRE13movingCL}. However, when there is wall superheating $\Delta T_{cl}>0$ at the contact line, evaporation occurs and $\theta_{V}>\theta_{micro}$, leading to an acceleration of the dewetting process \cite{JFM22}, which is consistent with experimental observations \cite{PRF16}.
A simplified approach that provides the dewetting speed $u_{d}$ as a function of $\Delta T_{cl}$ and $\theta_{micro}$ has been proposed \cite{EPL23}, which obviates the need for simulating a non-stationary dewetting problem. Through Eq.~\eqref{eq:Cox-Voinov}, the dependence of $\theta$ on $\Delta T_{cl}$ and $\theta_{micro}$ can also be determined. The characteristic length $L$ is known; it is proportional to the width of the dewetting ridge. Fig.~\ref{fig:FC72SBPHP} plots $u_{d}$ (left axis) and $\theta$ (right axis) as functions of $\Delta T_{cl}$; further details can be found in \cite{JFM22}. In summary, dewetting of liquid films is governed by nanoscale effects and local physical quantities such as $\theta_{micro}$ and the wall superheating $\Delta T_{cl}$ at the contact line position.

\subsubsection{Dynamic wetting transition}\label{BareMen}

\begin{figure}[ht]
  \centering
  \includegraphics[width=7cm,clip]{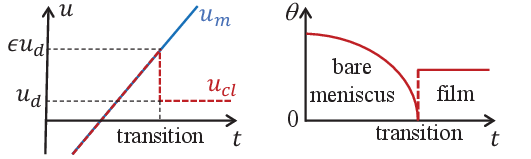}
  \caption{Schematic of the dynamic wetting transition according to \citet{Gao19}.}\label{fig:DynTrans}
\end{figure}

The liquid film is absent under certain conditions discussed below, and the meniscus shape is nearly spherical, as depicted in Fig.\ref{fig:dotm_menisci}b. This represents the second possible interfacial geometry, the bare meniscus, where the contact line moves at the same velocity $u_{m}$ as the meniscus. When it recedes fast enough, film deposition occurs, which leads to the interfacial geometry of Fig.\ref{fig:dotm_menisci}a. The onset of film deposition is called the dynamic wetting transition.
Studies, both theoretical and experimental \cite{Snoeijer06a}, have demonstrated that the liquid film is deposited when the velocity $u_{m}$ exceeds a certain threshold value, which is greater than the dewetting speed $u_{d}$ discussed previously \cite{Gao19}.
As the gap between the threshold and $u_{d}$ grows \cite{Gao19} with the contact angle (like $u_{d}$), we assume that the gap size $\propto u_{d}$. The threshold velocity can thus be defined as $\epsilon u_{d}$ by introducing a constant factor $\epsilon> 1$. Fig.~\ref{fig:DynTrans} illustrates the dynamic wetting transition, where the meniscus and contact line velocities are schematized as functions of time. Prior to the transition, $u_{m}<\epsilon u_{d}$ (indicated by a negative value representing the meniscus advancing the dry tube wall), and the contact line moves together with the meniscus, $u_{cl}=u_{m}$. Once the threshold is attained, $\theta=0$, and the contact line speed $u_{cl}$ suddenly drops to $u_{d}$. In the calculation below, we have chosen $\epsilon=2$, which is consistent with the numerical simulations of \citet{Gao19}.

In contrast to the film case discussed earlier, where $u_{cl}$ is directly determined by local quantities such as $\theta_{micro}$ and $\Delta T_{cl}$, $u_{cl}=u_m$ is imposed in this geometry. Here, $\theta$ is governed by Eq.~\eqref{eq:Cox-Voinov}, where $Ca=\mu u_{m}/\sigma$. Only the curvature of the meniscus, which may deviate from a circular shape, is known to be related to $L$. As $\theta=0$ at the transition, Eq.~\eqref{eq:Cox-Voinov} can be employed to determine $L$ for the bare meniscus case,
\begin{equation}\label{eq:Cox-Voinov2}
  L={\ell_{V}}\exp\left(\frac{\theta_{V}^3\sigma}{9 \mu u_{d}\epsilon}\right),
\end{equation}
where $u_{d}$ is the dewetting speed, which is determined by $\theta_{micro}$ and $\Delta T_{cl}$ in Fig.~\ref{fig:FC72SBPHP}.

\subsubsection{Evaporation/condensation for the bare meniscus case}\label{MassExMen}

Phase change over the interface should be considered separately for the two above geometries. Consider the total mass flux from bare meniscus. In all previous PHP models, it was introduced phenomenologically and quite arbitrarily. This value is however crucial for the evaluation of oscillation start-up \cite{IJHMT16}. Here we propose a physical approach to its evaluation.

Since the horizontal meniscus width is small with respect to the PHP size, the superheating can be assumed spatially homogeneous throughout the meniscus width and equal to $\Delta T_{cl}$. The 2D interface profile can be approximated as circular (Fig.~\ref{fig:dotm_menisci}b) because the inertial effects are weak in the small-diameter tubes. This circle has a radius of curvature $r_{m}=r/\cos\theta$ ($r=d/2$ is the tube radius) and forms the apparent contact angle $\theta$ with the wall. The local liquid thickness $\delta$ is linked to the local interface slope $\phi$ as
\begin{equation}\label{delta}
\delta(\phi)=r(1-\cos \phi/\cos\theta),
\end{equation}
that monotonously grows from 0 to $r$ while $\phi$ varies from $\theta$ to $\pi/2$.

As discussed in \autoref{sec:Jcl}, when calculating the total mass exchange in the contact line region, the local mass flux $j$ can be approximated by Eq.~\eqref{eq:jR}. Therefore, by using the above geometrical relation in Eq.~\eqref{zeta} and substituting $\zeta$ into Eq.~\eqref{eq:jR} one obtains
\begin{equation}\label{eq:j(s)m}
   j(\phi)= \frac{k_{l}\Delta T_{cl}}{r\cal L}\frac{{\sin \phi \cos \theta }}{{\alpha \sin \phi \cos \theta + \phi \left( {\cos \theta - \cos \phi } \right)}},
\end{equation}
where $\alpha\equiv R^\textrm{i} k_{l}/r$.

Let us evaluate the mass evaporation rate $J$ over the part of axially symmetric interface adjacent to the contact line until a point given by the angle $\phi\in(\theta,\pi/2)$ (Fig.~\ref{fig:dotm_menisci}b). Because of the rotational symmetry, the surface integral reduces to angular:
 \begin{equation}\label{eq:Jcl(phi)}
J(\phi)=2\pi \int_\theta^{\phi} (r-\delta(\varphi)) j(\varphi) r_{m}\mathrm{d}\varphi =\frac{\pi d k_{l} \Delta T_{cl}}{{\cal L}} w(\phi),
\end{equation}
where
\begin{equation}\label{eq:W(phi)}
w(\phi)= \frac{1}{\cos\theta}\int_{\theta}^{\phi} \frac{\sin\varphi \cos\varphi}{\alpha \sin\varphi \cos\theta+ (\cos\theta-\cos\varphi)\varphi} \mathrm{d}\varphi.
\end{equation}
Note that $\phi=\pi/2$ corresponds to the mass exchange rate over the entire interface,
\begin{equation}\label{eq:JclJm}
J(\pi/2)=\frac{\pi d k_{l} \Delta T_{cl}}{{\cal L}}W,
\end{equation}
where
\begin{equation}\label{eq:W}
W= \frac{1}{\cos\theta}\int_{\theta}^{\pi/2} \frac{\sin\varphi \cos\varphi}{\alpha \sin\varphi \cos\theta+ (\cos\theta-\cos\varphi)\varphi} \mathrm{d}\varphi
\end{equation}
is a function of $\alpha$ and $\theta$.
\begin{figure}[ht]
\centering
\includegraphics[height=4.5cm,clip]{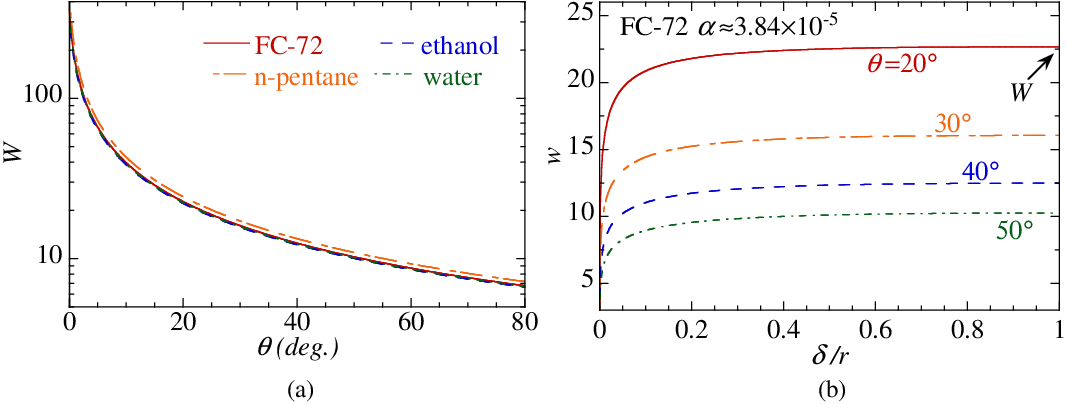}
\caption{Behavior of the function $w$. (a) Dependence of $W\equiv w(\phi=\pi/2)$ on $\theta$ for different $\alpha$. (b) Dependence of $w$ on $\delta$ for FC72 with $\alpha\approx 3.84\times 10^{-5}$ for different $\theta$.}\label{fig:W}
\end{figure}
For $r=\SI{1}{mm}$, one calculates $\alpha\approx 2.15\times 10^{-5}$ for n-pentane, $\alpha\approx 4.64\times 10^{-5}$ for ethanol, and $\alpha\approx 4.33\times 10^{-5}$ for water at saturation of \SI{1}{bar}. The relationship between $W$ and $\theta$ can now be computed for these fluids (as shown in Fig.~\ref{fig:W}a); $W$ is strongly influenced by $\theta$ but is almost independent of the fluid.

The relative contribution of the contact line region $J_{cl}$ and the meniscus central part $J_m$ can be analyzed by varying $\phi$ or, equivalently, $\delta$ related to $\phi$ via Eq.~\eqref{delta}. Obviously, $J_{cl}$ can be associated with $J(\phi)$ given by Eq.~\eqref{eq:Jcl(phi)} while $$J_{m}=J(\pi/2)-J(\phi)\equiv \frac{\pi d k_{l} \Delta T_{cl}}{{\cal L}}[W-w(\phi)].$$
Examples of $w(\delta)$ variation are plotted in Fig.~\ref{fig:W}b. With growing $\delta$, $w$ rapidly saturates to $W$, which signifies that the flux $J_{cl}$ from the contact line vicinity is by far dominant, and $J_{m}$ is much smaller. Therefore, for simplicity, one can attribute the evaporation flux $J(\pi/2)$ from the entire interface to the contact line vicinity.

To summarize, one can consider that the total mass exchange from the bare meniscus is given by the contact line region flux
\begin{equation}\label{eq:Jclbare}
J_{cl}=\frac{\pi d k_{l} \Delta T_{cl}}{{\cal L}}W,
\end{equation}
where $W$ depends on $\theta$ according to Eq.~\eqref{eq:W}. Note that for the meniscus geometry without the film, $\theta$ is a function of both $u_{cl}\equiv u_m$ and $\Delta T_{cl}$ as given by Eq.~\eqref{eq:Cox-Voinov}, so $J_{cl}$ depends on both $u_{cl}$ and $\Delta T_{cl}$.

\subsubsection{Evaporation/condensation for the deposited film case}

One needs to establish now the total mass flux for the deposited film case. As established above, the mass flux $J_m$ from the central meniscus part is negligible; one can thus consider only two remaining contributions (cf. Fig.~\ref{fig:dotm_menisci}a): the flat film contribution $J_f$ and the contact line region contribution $J_{cl}$. Let us start with this latter term.

\begin{figure}[ht]
  \centering
  \includegraphics[height=4cm,clip]{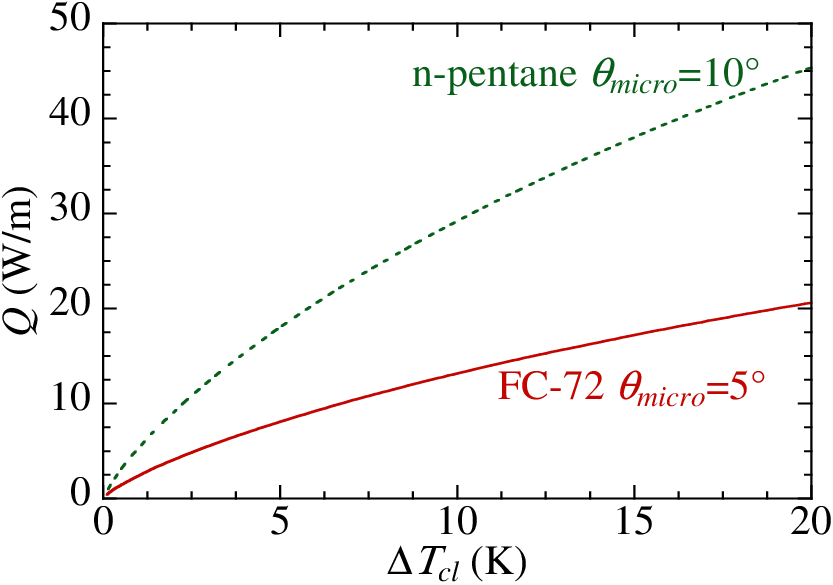}
  \caption{The value of $Q$ as a function of $\Delta T_{cl}$ for FC-72 with $\theta_{micro}=5^\circ$ (red solid curve) and for n-pentane with $\theta_{micro}=10^\circ$ (green dashed curve) for the conditions of  \autoref{tab:expsum}.}\label{fig:Q}
\end{figure}

The contact line motion (receding) appears because of the film dewetting phenomenon accelerated by evaporation \cite{JFM22,EPL23}. As discussed in sec.~\ref{MenDep}, for the dewetting case, both $u_{cl}\equiv u_d$ and $\theta$ are functions of $\Delta T_{cl}$. The contact line region contribution $J_{cl}$ can be approximated with the same expression \eqref{eq:Jclbare} as for the bare meniscus in spite of different curvatures in these cases. This is possible because of the strong localization of phase change at the contact line vicinity (Fig.~\ref{fig:W}b) where the slope is defined mainly by $\theta$. By using in Eq.~\eqref{eq:Jclbare} (more precisely, in $W$) the value of $\theta$ from Fig.~\ref{fig:FC72SBPHP}, one obtains
\begin{equation}\label{eq:Jcl}
J_{cl}=\frac{\pi d Q}{{\cal L}},
\end{equation}
where the heat exchange rate $Q= k_{l} \Delta T_{cl}W$ per unit contact line length is a function of only $\Delta T_{cl}$. Note the difference from the meniscus case, where $u_{cl}=u_m$ varies independently of $\Delta T_{cl}$, and $J_{cl}$ is a function of both these quantities. Fig.~\ref{fig:Q} shows the variation of $Q$ with $\Delta T_{cl}$ for two fluids used in the subsequent PHP simulations.

Consider now the mass exchange at the film interface. The mass flux $J_f$ can be obtained by integrating Eq.~\eqref{eq:jR} over the area of the film of a uniform thickness $\delta$. Since $\delta\gg\SI{100}{nm}$, the interfacial resistance can be safely neglected, and $J_f$ is an integral of $j$ from the contact line to the meniscus at $x=x_m$,
\begin{equation}\label{eq:Jf}
  J_{f}= U_{f} \frac{\pi d_{f} }{{\cal L}} \int_{x_{cl}}^{x_{m}} \left[ {{T_{w}}(x) - {T_{sat}}} \right]\mathrm{d}x.
\end{equation}
Here $U_{f}= \varsigma k_{l}/ \delta $ is the heat exchange coefficient of the film and $d_{f}=d-2\delta$. The correction factor $\varsigma$ first introduced in \cite{IJHMT10} accounts for the spatial variation \cite{JFM21} of the film thickness, cf. Appendix~\ref{app:form}.

Note that $J_{cl}$ is usually negligibly small in comparison with $J_{f}$ when the film is present, cf.~Appendix \ref{app:rel}, like for the stable oscillation regime considered hereafter. For this reason we neglect the $J_{cl}$ time variation caused by the dependence of $\theta$ on $u_{cl}$ and $\Delta T_{cl}$ (through $\theta_V$ in Eq.~\eqref{eq:Cox-Voinov}). We use a fixed value of $W$ (cf. Table~\ref{tab:expsum}) that corresponds to $\theta\approx 20^\circ$ (Fig.~\ref{fig:W}a) conforming to typical parameters applied in our simulation discussed in sec.~\ref{ResSec}. However, $J_{cl}$ is essential when the film is absent \cite{IJHMT16}, in particular at the PHP start-up at the beginning of functioning or after a temporary stop-over, which is a common scenario of the intermittent functioning regime \cite{MST19}. For these studies (which are yet to be performed), the impact of $\theta$ variation on $J_{cl}$ is important and should be accounted for in a more rigorous way.

\section{Model formulation}\label{modelsec}

\begin{figure}[htb]
  \centering
  \includegraphics[width=5cm,clip]{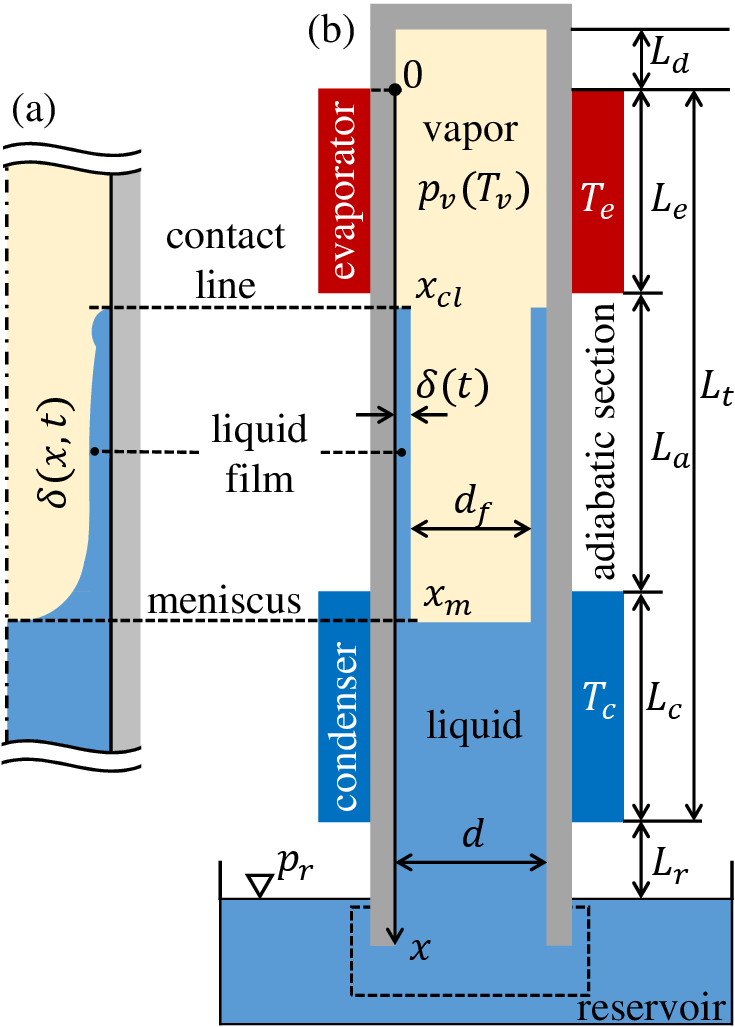}
  \caption{(a) Curved vapor-liquid interface in a capillary and (b) lumped geometry representation of the vapor-liquid interface for a single-branch PHP.}\label{fig:SBPHP}
\end{figure}

One can now formulate the model, which includes the previously discussed features. Figure~\ref{fig:SBPHP}b depicts a schematic of a single-branch PHP, which comprises a capillary tube, an evaporator with length $L_{e}$, a condenser with length $L_{c}$, and an adiabatic section with length $L_{a}$. One end of the capillary tube is sealed while the other is open and submerged in a reservoir of constant pressure $p_{r}$. The capillary tube length from the condenser to the liquid level in the reservoir is denoted by $L_{r}$. In several experimental setups, additional dead space between the sealed end and the evaporator is required for pressure and temperature sensors, in which the vapor remains stagnant. An equivalent length $L_{d}$ represents this space.
The main assumptions concerning the analysis are summarized as follows:
\begin{enumerate}
  \item One-dimensional plug/slug flow regime;
  \item Actual curved liquid-vapor interface (Fig.\ref{fig:SBPHP}a) is described with the lumped geometry, in which the meniscus is flat and the effective thickness $\delta(t)$ of liquid film is uniform but varies with time.
  \item Spatially averaged parameters such as liquid velocity and vapor pressure are used;
  \item Vapor bulk obeys the ideal gas law; the temperature of the free interface is the saturation temperature corresponding to the vapor pressure, so there exists a thin thermal boundary layer near its surface;
 \item Free contact line of liquid film recedes conforming to the physics of dewetting discussed above.\label{ass4}
\end{enumerate}

\subsection{Liquid plug}
Liquid plug description is similar to the superheated vapor and FEC models. The liquid is assumed incompressible and of constant density. The momentum equation reads \cite{IJHMT10}
\begin{equation}\label{eq:dot_mlVl}
  \frac{\mathrm{d}}{\mathrm{d}t} \left[(m_{l}+m_{l,i}) u_{l}\right] = (p_{v}-p_{r})S- F + m_{l}g,
\end{equation}
where the liquid plug mass is $m_{l}=\rho_{l}(L_{t}-x_{m}+L_{r})S$, $S=\pi r^2$ and $L_{t}=L_{e}+L_{a}+L_{c}$. $u_{l}$ is the plug velocity (of its center of mass), positive for the plug moving downwards. The added liquid mass $m_{l,i}=\rho_{l}L_{i}S$ in the reservoir (the region enclosed by the dashed line in Fig.~\ref{fig:SBPHP}b) participates in the oscillating motion. It is designated by an equivalent length $L_{i}$. The precise value of $L_{i}$ is hard to be quantified experimentally. Therefore, in simulations, $L_{i}$ serves to slightly adjust the oscillation period to agree with the experiment \cite{IJHMT10}.

On the right-hand side of Eq.~\eqref{eq:dot_mlVl}, the last term is the gravity force with the acceleration $g$, while the first represents the pressure force. The second term corresponds to the pressure loss, whose sign is opposite to the plug velocity. It consists of two parts: viscous (regular) pressure loss (i.e. the friction force) and singular pressure loss appearing at the open tube end,
\begin{equation}\label{eq:Ff}
F = \left[ K \pi d (L_{t}-x_{m}+L_{r}+L_{f}) +bS \right] \frac{\rho_{l} u_{l}^2}{2} \mathrm{sgn}(u_{l}).
\end{equation}
The viscous friction is calculated by the first term in the square brackets, in which $K$ is the Fanning friction factor \cite{IJHMT10},
\begin{subequations}\label{eq:FanningFric}
\begin{numcases}{K=}
  0, & if $Re_{l}=0$ \\
  16/Re_{l}, & if $0< Re_{l} < 2100$,  \\
  0.0791Re_{l}^{-0.25}, & otherwise ,
\end{numcases}
\end{subequations}
where the second line is for laminar flow and the last line is the turbulent flow described by Blasius correlation applicable for $Re_{l} \leq 10^5$ (the condition is largely satisfied in PHP). The singular term including $b$ is discussed below. In this description, the viscous friction is assumed to be equal to that of a fully developed flow in tubes. However, Eq.~\eqref{eq:FanningFric} underestimates the liquid pressure loss in PHP. Oscillatory motion creates an additional viscous dissipation in the reservoir. To compensate for this effect, an additional length $L_{f}$ is introduced into the friction force; its value however is hard to determine precisely. Therefore, $L_{f}$ is tuned to accord with the experiment, mainly for the fine adjustment of the meniscus oscillation amplitude.

Near the tube outlet in the reservoir, a sudden change in the flow area exerts the singular pressure loss \cite{Quere99}. It is represented by the flow resistance coefficient $b$ in Eq.~\eqref{eq:Ff},
\begin{subequations}\label{eq:b}
\begin{numcases}{b=}
  0.5, & if $u_{l} > 0$ \\
  0.25, & if $u_{l}\leq 0$.
\end{numcases}
\end{subequations}

\subsection{Vapor phase}\label{secvap}

The vapor description is similar to the superheated vapor and FEC models. The bulk of the vapor phase in the single-branch PHP simulation is described by its temperature $T_{v}$ and pressure $p_{v}$. Previous studies \cite{ATE21} have demonstrated that the vapor in the single-branch PHP is mostly at the superheated state; therefore, the vapor behavior can be well approximated by the ideal gas law:
\begin{equation}\label{eq:EOS}
  p_{v}=\frac{m_{v}R_{v}T_{v}}{\Omega_{v}}.
\end{equation}
where $m_{v}$ and $\Omega_{v}=Sx_m$ are the mass and the volume of vapor, respectively.
Equation~\eqref{eq:EOS} provides the elastic response for oscillation, which acts as the restoring force in vapor compression and expansion.

The temperature of the liquid-vapor interface is assumed to be equal to the saturation temperature $T_{sat}$ corresponding to $p_{v}$, while $T_{v}$ is determined by the energy balance \cite{shafii1,Comment11}:
\begin{equation}\label{eq:dot_Tv}
  m_{v} c_{v} \dot{T_{v}} = \dot m_{v} R_{v}T + U_{v} \pi d \int_{0}^{x_{cl}} \left[ T_{w}(x) - T_{v} \right]\mathrm{d}x - p_{v}\dot\Omega_{v}
\end{equation}
On the right-hand side, the second term represents the sensible heat exchange between the vapor and dry tube wall, where $U_{v}$ is the heat exchange coefficient. It is usually a small value since  $U_{v}=k_{v}Nu_{v}/d$ with the Nusselt number $Nu_{v}\approx 6$ \cite{Gully14}. This term is unsubstantial compared to the latent heat exchange through the liquid film, but is essential for the PHP startup \cite{IJHMT16}, where the film is absent.

Because of the weak heat diffusion in the vapor phase, a thermal boundary layer exists, and the temperature $T_{v}$ of the vapor bulk may differ from the free interface temperature $T_{sat}$, which is normally several degrees lower than $T_{v}$ as observed experimentally \cite{Gully14}.

The vapor mass change $\dot m_{v}$ is the sum of mass fluxes over the interface,
\begin{equation}\label{eq:dot_mv}
  \dot m_{v}= J_{cl}+J_f
\end{equation}
where $J_f$ is defined by Eq.~\eqref{eq:Jf}, and $J_{cl}$ is defined by either Eq.~\eqref{eq:Jcl} or Eq.~\eqref{eq:Jclbare} depending on the film existence as discussed in \autoref{secrec}. The value of $\varsigma$ in $J_f$ (Appendix \ref{app:form}) is used to fine-tune the oscillation amplitude to conform to the experiment, $\varsigma \approx 1.5$.

If one neglects the pressure losses, the heat and mass transfers, Eqs.~(\ref{eq:dot_mlVl},\ref{eq:EOS},\ref{eq:dot_Tv}) result in the eigenfrequency of oscillations
\begin{equation}\label{omega0}
 \omega_0=\sqrt{\frac{S}{\bar m_{l}+m_{l,i}}\left(\frac{\gamma p_{r}S}{\bar \Omega_{v}}+g\rho_l\right)},
\end{equation}
where $\gamma=1+R_v/c_v$ is the adiabatic index, and a bar signifies a quantity averaged over the oscillation period. The first term in the brackets corresponds to the adiabatic vapor compressibility \cite{IJHMT10}, while the second is the gravity effect \cite{Rao13}. Eq.~\eqref{omega0} means that the frequency should increase with the reservoir pressure and decrease with both the liquid and vapor volumes in the capillary. Note that while the frequency of self-sustained oscillation remains close to $ \omega_0$, a (usually small) deviation caused by the nonlinearity necessarily occurs.

\subsection{Liquid film}
Two quantities characterize the liquid film: its mass $m_{f}$ and its length ${L}_{lf}$, which is a distance between $x_{cl}$ and $x_{m}$. Their variations with time are described in the OFT model. All three quantities are linked by the equality
\begin{equation}\label{eq:mf}
  m_{f}= S_{f} (x_{m}-x_{cl})\rho_{l},
\end{equation}
where
\begin{equation}\label{Sf}
S_{f}=\pi[ r^2-(r-\delta)^2]\equiv\pi \delta (d-\delta)
\end{equation}
is the film cross-section area. The effective (spatially averaged, cf. Appendix \ref{app:form}) film thickness $\delta$ can be deduced from the above equations as
\begin{equation}\label{eq:delta}
 \delta = r-\sqrt{r^2-\frac{m_{f}}{ \pi \rho_{l}(x_{m}-x_{cl}) }}.
\end{equation}

\subsubsection{Film length variation}
The film length ${L}_{lf}={x}_{m}-{x}_{cl}$ varies because of the meniscus oscillation and the contact line displacement:
\begin{equation}\label{eq:dotLf}
\dot{L}_{lf} = u_{m}-u_{cl}.
\end{equation}

As the liquid plug deposits a film, the meniscus moves towards the center of the plug due to the mass loss, resulting in a slightly higher $u_{m}$ compared to $u_l$. However, in the single-branch PHP, the mass of liquid plug is significantly larger than that of the liquid films. Hence, the difference between $u_{m}$ and $u_l$ can be safely ignored,
\begin{equation}\label{eq:u_m}
  \dot x_{m}\equiv u_m = u_{l}.
\end{equation}

According to assumption \ref{ass4}, the OFT model uses the physics of the contact line motion discussed in sec.~\ref{secrec}. The contact line velocity is described as
\begin{subequations}\label{eq:dot_xcl}
\begin{numcases}{u_{cl} =}
  u_{m} & if $x_{m} = x_{cl}$ and $u_{m} \leq \epsilon u_{d}$, \label{eq:dot_xcl1}\\
  %\min \{{u_{m}, \epsilon u_{d}}\} & if $x_{m} = x_{cl}$ and $u_{m} >0$, \label{eq:dot_xcl2}\\
  u_{d}  & otherwise.  \label{eq:dot_xcl2}
\end{numcases}
\end{subequations}
The second line indicates that the contact line is the edge of a liquid film (Fig.~\ref{fig:dotm_menisci}a), which recedes at the dewetting speed $u_{d}= f\left( \Delta T_{cl}, \theta_{micro} \right)$. It is assumed here that for $\Delta T_{cl}<0$, $u_{d}= 0$. Eq.~\eqref{eq:dot_xcl1} is written for the ``bare'' meniscus case (Fig.~\ref{fig:dotm_menisci}b) that occurs in two situations: (i) meniscus advancing over the dry tube wall ($u_{m} < 0$); (ii) meniscus receding with speed lower than the threshold $\epsilon u_{d}$ of the dynamic wetting transition.

\subsubsection{Film mass variation}

Liquid film exchanges mass with the liquid plug and with the vapor bubble. The film mass variation reads
\begin{equation}\label{eq:dot_mf}
\dot m_{f}=\dot m_{dep}-J_{f}-J_{cl},
\end{equation}
where $J_f$ is given by Eq.~\eqref{eq:Jf}, and $J_{cl}$, by Eq.~\eqref{eq:Jcl}. Consider now the film deposition rate $\dot m_{dep}$. It is expressed as
\begin{subequations}\label{eq:dot_mdep}
\begin{numcases}{\dot m_{dep} =}
  0, & $x_{cl} = x_{m}$ and $u_{m} \leq \epsilon u_{d}$, \label{eq:dot_mdep1}\\
  S_{f} u_{m} \rho_{l}, & $x_{cl} < x_{m}$ and $u_{m} < 0$, \label{eq:dot_mdep2}\\
 S_{dep}u_{m} \rho_{l},  & otherwise, \label{eq:dot_mdep3}
\end{numcases}
\end{subequations}
where $S_{dep}=\pi \delta_{dep} (d-\delta_{dep})$ is the cross-section area of the deposited film that has the thickness $\delta_{dep}$ defined by Eq.~\eqref{eq:delta_dep}. Equation~\eqref{eq:dot_mdep} may be regarded as a mass exchange rate between the liquid film and the plug. Equation~\eqref{eq:dot_mdep1} corresponds to the situation where the contact line coincides with the meniscus (Fig.~\ref{fig:dotm_menisci}b). This occurs when the plug advances over the dry tube ($u_{m}<0$), or the plug recedes ($u_{m} > 0$) slower than the threshold of film deposition. Usually, $u_{m}$ is large enough, so a plug receding without film deposition rarely occurs in PHPs. Equation~\eqref{eq:dot_mdep2} corresponds to the situation where the plug advances over an existing film ($x_{cl}<x_{m}$); the plug absorbs the film. Since $u_{m}<0$, $\dot m_{dep}$ is negative in this case.
Equation~\eqref{eq:dot_mdep3} describes two possible film deposition scenarios: (i) $x_{cl}=x_{m}$ and $u_{cl} < u_{m}$ (complementary to the clause~\eqref{eq:dot_mdep1}): the contact line coincides with the meniscus, and the meniscus recedes faster than the contact line. Hence, a new film is about to be deposited. (ii) $x_{cl} < x_{m}$ and $u_{m}\geq 0 $ (complementary to the clause~\eqref{eq:dot_mdep2}); the liquid film exists and keeps to be deposited.

\subsection{Boundary conditions}
In the present calculations, the inner wall temperature in the evaporator section $T_{e}$ and in the condenser section $T_{c}$ are assumed to be uniform and invariable with time. These conditions are achievable in many experiments where both heating and cooling are efficient enough to impose the constant temperature to the tube \cite{IJHMT10,Rao15,Rao17}. The adiabatic section assumes a linear variation from $T_{e}$ to $T_{c}$.
A set of six ordinary differential equations (\ref{eq:dot_mlVl}, \ref{eq:dot_Tv}, \ref{eq:dot_mv}, \ref{eq:u_m}--\ref{eq:dot_mf}) defines the single-branch PHP behavior. As an analytical solution is impossible to be found, the equations are solved numerically with the 4th order Runge-Kutta method.

\section{Asymptotic results}\label{asympRes}

\subsection{Behavior at the film deposition instant}

Note that at the moment of film deposition, the effective film thickness $\delta$ does not necessarily coincide with $\delta_{dep}$ in spite of the enforced condition \eqref{eq:dot_mdep3}. This paradox can be explained as follows. Consider the time-lapse $\Delta t$ at the beginning of the film deposition. According to Eq.~\eqref{eq:dot_xcl}, $u_{m}=\epsilon u_{d}$ and $u_{cl}= u_{d}$ at this instant. Since the film length is small, $\dot m_{v}$ contribution to $\dot m_{f}$ is negligible, so $\dot m_{f}\approx\dot m_{dep}$ according to Eq.~\eqref{eq:dot_mf}. From the condition \eqref{eq:dot_mdep3},
$$m_{f}\approx\pi \delta_{dep} (d-\delta_{dep})\epsilon u_{d} \rho_{l}\Delta t.$$
On the other hand, from Eqs. (\ref{eq:mf}--\ref{Sf}),
$$m_{f}\approx \pi \delta (d-\delta) (x_{m}-x_{cl})\rho_{l}\approx \pi \delta (d-\delta)u_{d}(\epsilon-1)\rho_{l}\Delta t.$$

Equalizing these two expressions under the assumptions  $\delta,\,\delta_{dep} \ll d$ results in

\begin{equation}\label{eq:delta0}
\delta\approx\frac{\epsilon}{\epsilon-1} \delta_{dep},
\end{equation}
so one can see that $\delta$ and $\delta_{dep}$ are indeed different. This occurs because the contact line moves simultaneously with the meniscus receding. The choice of $\epsilon$ affects the very initial stage of film evolution and does not have a notable impact on the overall heat exchange. Note that without the plummet in $u_{cl}$ (i.e. assuming $\epsilon=1$), $\delta$ would be infinite according to Eq.~\eqref{eq:delta0}.

\subsection{Film removal condition}
A singularity appears in Eq.~\eqref{eq:delta} when the meniscus approaches the contact line $x_{m}\to x_{cl}$, cf. Appendix \ref{appAsympApp}. In reality, the capillary action prevents such a singularity. A short liquid film merges with the plug, and the interface rapidly restores to a smooth meniscus \cite{JFM21}. To avoid this singularity, which is an artifact of the 1D model, a film removal condition is introduced: when the film length $|x_{m}-x_{cl}| < \delta$ and $u_{m}<u_{cl}$, the liquid film is removed from the system, and its mass is reassigned to the adjacent liquid plug. %In Fig.~\ref{fig:Rao15}b below, the condition is satisfied near the end of each period.

\section{Results and discussion}\label{ResSec}
\subsection{Comparisons with the FEC model}\label{ResFEC}

Unlike the OFT model, the FEC model requires a value of $\delta$ as a constant input parameter. However it should be consistent with the meniscus velocity. One can determine $\delta$ with an iteration algorithm \cite{IarATE17}. The first iteration of the simulation adopts a practical value as its input, for instance, $\sim\SI{100}{\micro m}$. Then the output results are used to find the root mean square plug velocity $u_{m,RMS}$  over an oscillation period. Eq.~\eqref{eq:delta_dep} can now be used to predict the thickness $\delta'$ corresponding to $u_{m,RMS}$. $\delta'$ is used as an input for the next iteration. This process is repeated until an equality (within a given accuracy, typically $<1\%$) between $\delta$ and $\delta'$ is achieved.

Instead of Eq. \eqref{eq:dot_xcl} in OFT model, the FEC model uses the film mass conservation to describe the film edge receding
\begin{equation}
u_{cl} =\begin{cases}
u_{m} & \mbox{ if } x_{m}=x_{cl},\; u_{m}<0,\\
J_{f,e}/(\rho_l\pi d\delta) & \mbox{ otherwise},
\end{cases}\label{dxf}
\end{equation}
where $J_{f,e}\geq 0$ is a part of $J_f$ corresponding to the film portion that situates in the evaporator \cite{IJHMT16}.

To understand the difference between the OFT and FEC results, we simulate the single-branch PHP geometry of \citet{IJHMT10} for the same parameters (Table~\ref{tab:expsum}).
\begin{figure}[htb]
  \centering
  \includegraphics[width=7cm,clip]{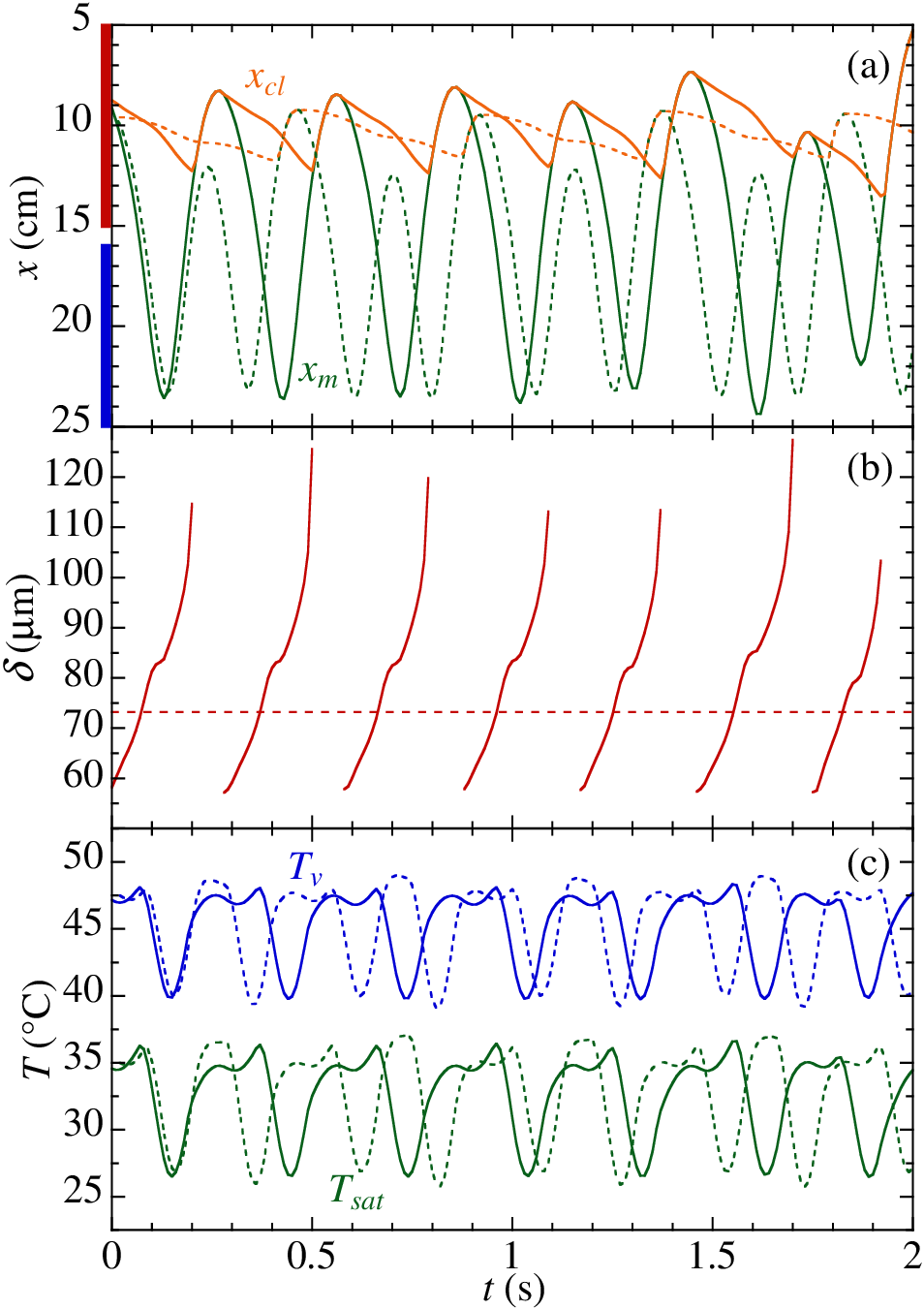}
  \caption{Comparison between the OFT (solid curves) and FEC (dashed curves) models for the PHP geometry of \citet{IJHMT10}. (a) Meniscus ($x_{m}$) and  contact line ($x_{cl}$) displacements; the locations of the evaporator and condenser are shown with red and blue bars, respectively. (b) Effective film thickness $\delta$; (c) Vapor temperature $T_{v}$  and saturation temperature $T_{sat}$ corresponding to the vapor pressure.}
  \label{fig:Comp_OFT_FEC}
\end{figure}
Figures~\ref{fig:Comp_OFT_FEC} show the temporal evolution of main PHP variables. The $L_f$ and $L_i$ values were fitted for the OFT model to achieve the oscillation period of \SI{0.28}{s} observed experimentally \cite{IJHMT10}. The same values were used for the FEC model. One can see that contrary to the OFT results, the FEC curves show the period doubling (a smaller amplitude of each second oscillation) that causes a higher frequency. The magnitude of oscillations is quite similar in both models. The absence of period doubling in the OFT model is related to the fast contact line receding.

Fig.~\ref{fig:Comp_OFT_FEC}b plots the temporal variation of the effective thickness. In the OFT model, at the onset of film deposition, $\delta$ starts growing from $\approx\SI{57.8}{\micro m}$, which corresponds to the value predicted by Eq.~\eqref{eq:delta0} with the meniscus velocity $\approx\SI{0.37}{m/s}$ observed at this moment. The value $\delta_{FEC}\approx\SI{73.2}{\micro m}$ used in the FEC model (a dashed horizontal line in Fig.~\ref{fig:Comp_OFT_FEC}b) is a result of the above iteration scheme; it corresponds to $u_{m, RMS}\approx\SI{1.23}{m/s}$.
$\delta_{FEC}$ compares well to the value $\bar\delta_{OFT}\approx \SI{79}{\micro m}$ obtained by the time averaging of the film thickness variation in the OFT model.

These results show the viability of the FEC model, at least for this particular case. One notes however that the OFT model is expected to be more precise for the case of the multi-branch PHP, where numerous liquid films of \emph{a priori} different thicknesses exist. In addition, the OFT model provides the film thickness within a single calculation while multiple calculations are required for iterations of the FEC model.

\begin{table}[tbp]
\centering
\begin{tabular}{cccc}\toprule
                    & With FEC     & With experiment & Wetting study  \\ \hline
tube orientation    & horizontal   & vertical                & horizontal      \\
working fluid       &  n-pentane   & FC-72                   &  water       \\
$d$ (mm)            &  2           & 2                       &  2              \\
$L_{e}$ (cm)        & 15           & 20                      & 15              \\
$L_{a}$  (cm)       & 1            & 1                       & 1               \\
$L_{c}$ (cm)        & 25           & 20                      & 25              \\
$L_{r}$ (cm)        & 10           & 20                      & 10              \\
$T_{e}$ ($^\circ$C) & 45           & 44                      & 70              \\
$T_{c}$ ($^\circ$C) & 10           & 16                      & 40              \\
$p_{r}$ (kPa)       & 90           & 50                      & 12.35           \\
$h_{r}$ (cm)        & 10           & --                      & 0              \\
$L_{d}$ (cm)        & 50           & 90                      & 50               \\
$L_{f}$ (cm)        & 30           & 30                      & 30              \\
$L_{i}$ (cm)        & 5            & 8                       & 5               \\
$\varsigma$         & 1.4          & 1.5                     & 1.4              \\
$\theta_{micro}$    & $10^\circ$   & $5^\circ$               & $10^\circ$, $20^\circ$, $30^\circ$       \\
$W$                 & 15           & 20                      & 15               \\ \bottomrule
\end{tabular}
\caption{Major experimental parameters and parameters used in the comparative numerical simulations. The $p_r$ value defines the saturation conditions for the determination of all fluid constants.}\label{tab:expsum}
\end{table}

\subsection{Experimental validation}

\citet{Rao15} have conducted single-branch PHP experiments using a vertical capillary tube and FC-72 as the working fluid. Their experiment setup is similar to that shown in Fig.~\ref{fig:SBPHP} with a circular transparent glass capillary tube. Two heat exchangers acting as evaporator and condenser impose constant temperatures (overall accuracy of $\pm 1^{\circ}$C) to the tube. The main PHP parameters are summarized in the last column of Table~\ref{tab:expsum}. For comparison with the OFT model, one can use the experimental data on the displacements (of the meniscus and the contact line), and the saturation temperature derived from the experimental vapor pressure.

\begin{figure}
  \centering
  \includegraphics[width=7cm,clip]{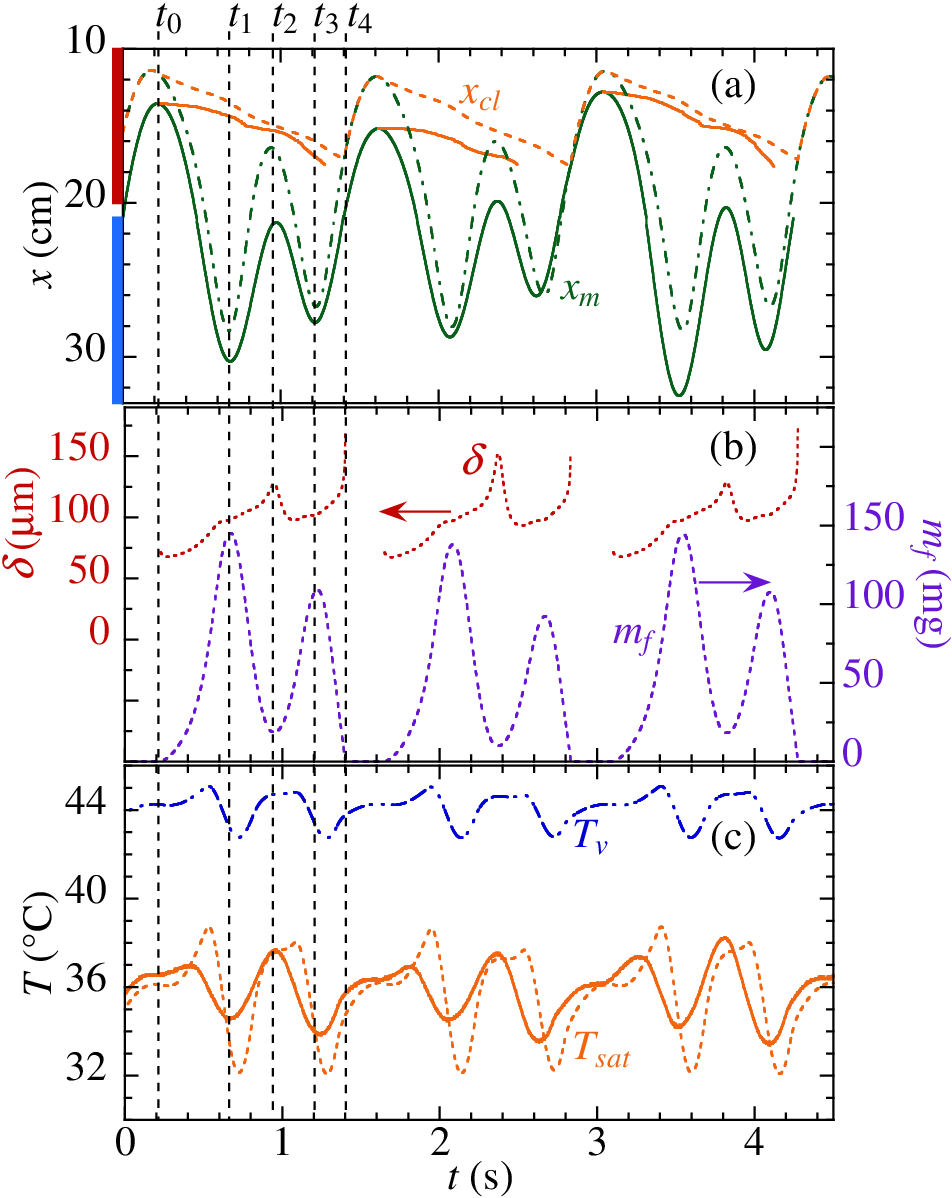}
  \caption{Comparison of the numerical results (dashed curves) and the experimental data (solid curves) of \citet{Rao15}. (a) Meniscus ($x_{m}$) and  contact line ($x_{cl}$) displacements; the locations of the evaporator and condenser are shown with red and blue bars, respectively. (b) Effective film thickness $\delta$ and film mass $m_{f}$; (c) Vapor temperature $T_{v}$  and saturation temperature $T_{sat}$ corresponding to the vapor pressure.}
  \label{fig:Rao15}
\end{figure}
\begin{figure*}[htb]
\centering
  \includegraphics[width=12.5cm,clip]{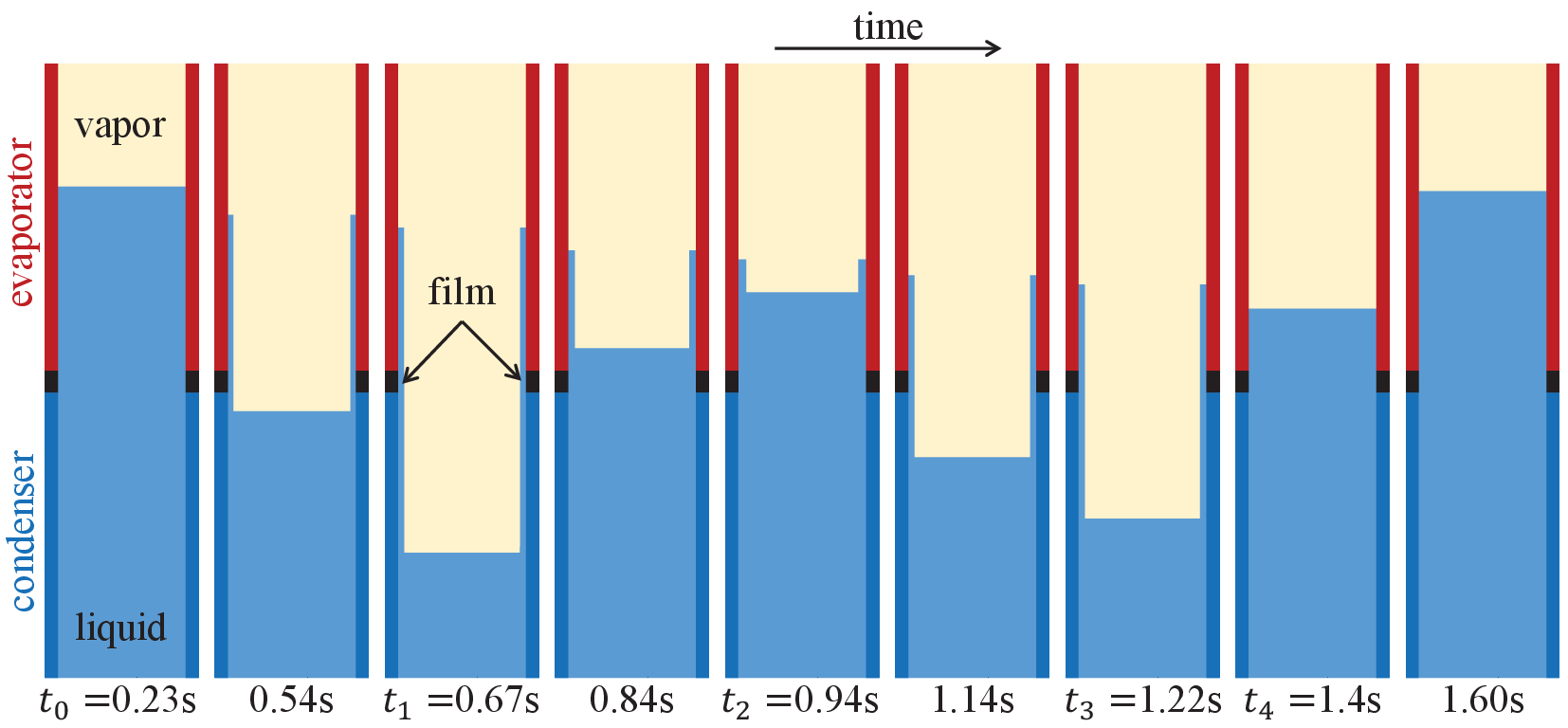}
  \caption{Model results showing the meniscus displacement and liquid film variation during an oscillation period, corresponding to the experimental setup of \citet{Rao15}. The corresponding supplementary movie is available.}
  \label{fig:SBPHPsnapshot}
\end{figure*}

In the original experiments of \citet{Rao15}, the evaporator temperature targeted by the thermal regulation was $46^{\circ}$C.  However, in their subsequent analysis \cite{Rao17} considering heat conduction of the glass tube, the temperature of the inner tube wall was estimated to be lower. \citet{Gully14} have revealed that the vapor temperature oscillates around $T_{e}$ in single-branch PHP. Therefore, in our calculation, the vapor temperature $\approx 44^{\circ}$C measured by \citet{Rao15} has been taken as the actual $T_{e}$ value.

Figure~\ref{fig:Rao15}a shows the temporal variation of $x_{m}$ and $x_{cl}$ over three oscillation periods. The numerical variation of liquid film thickness is plotted in Fig.~\ref{fig:Rao15}b. A comparison of the numerical and experimental saturation temperatures (the latter is deduced from the experimentally measured  vapor pressure) is shown in Fig.~\ref{fig:Rao15}c. As expected, the vapor temperature $T_{v}$ oscillates around $T_{e}$ and remains well above the saturation temperature.

Generally, a very good agreement between the experimental and simulation results is observed. The model well reproduces the behavior of major parameters, including the doubling of oscillation period and the shape of the $T_{sat}$ curves (Fig.~\ref{fig:Rao15}c). Remarkably, a good agreement is observed for the receding speed of the contact line. Note that there are no adjustable parameters in the film model. Unfortunately, the film thickness variation could not be compared as it was not measured experimentally.

\subsection{Film thickness evolution}

The simulated film thickness variation can be analyzed by using Figs.~\ref{fig:Rao15}. The corresponding vapor-liquid interface evolution is shown in Fig.~\ref{fig:SBPHPsnapshot}. The meniscus oscillates over all three tube sections, while the contact line always stays in the evaporator. The film deposition takes place at $t_0=\SI{0.23}{s}$, with the initial thickness $\SI{71}{\micro\meter}$ that agrees with Eq.~\eqref{eq:delta0}. Between $t_0$ and $t_1=\SI{0.67}{s}$, the meniscus keeps falling and depositing liquid, leading to an increase in $m_{f}$. At the end of the fall, the meniscus decelerates, and $\delta_{dep}$ (defined by Eq.~\eqref{eq:delta_dep}) decreases; the effective thickness $\delta$ does not. This occurs because the contact line receding leads to the dewetting ridge growth \cite{JFM22}, so the average thickness keeps growing. Its growth only decelerates before $t_1$ (Fig.~\ref{fig:Rao15}b). At $t_1$, the meniscus hits the trough and rises until $t_2=\SI{0.94}{s}$. The plug collects the film, and $m_{f}$ declines. Such an absorption does not lead to a $\delta$ variation, cf. Eqs.~(\ref{eq:delta}, \ref{eq:dot_mdep2}). However, between $t_1$ and $t_2$, $\delta$ increases because of the contact line receding, the same amount of liquid being averaged over shorter film length. This effect outperforms the film evaporation that tends to lessen $\delta$.
From $t_2$ to $t_3=\SI{1.22}{s}$, the meniscus falls and deposits liquid, leading to a growth in $m_f$. Regardless, $\delta$ decreases at the beginning of the meniscus fall because the deposited thickness $\delta_{dep}$ is smaller than the current thickness $\delta$. Immediately before the meniscus draws back to the contact line at $t_4=\SI{1.4}{s}$, the value of $\delta$ soars up in agreement with the analysis in Appendix~\ref{appAsympApp} before attaining the film removal condition. The average value of $\delta$ is $\sim$\SI{97}{\micro\meter} over a period.

\subsection{Impact of wetting properties}
It is known from experiment \cite{Hao14} that the wetting properties strongly impact the PHP performance; it decreases with the deterioration of wettability. To the best knowledge of the authors, the OFT model is the only one that accounts for this effect.
\begin{figure}[hbt]
  \centering
  \includegraphics[width=7cm,clip]{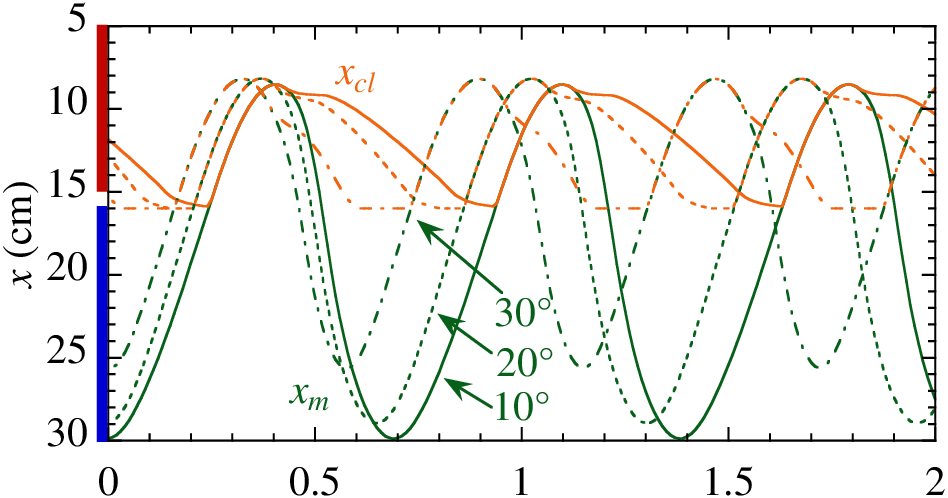}
  \caption{Dependence of the meniscus and contact line oscillation on the wetting properties for water. The parameter of the curves is $\theta_{micro}$.}
  \label{fig:OFT_water}
\end{figure}
The simulated within OFT model SBPHP is of the same geometry as that of \citet{IJHMT10}. The parameters used in the simulation are listed in Table~\ref{tab:expsum}, in the ``Wetting study'' column.

The PHP is simulated for different $\theta_{micro}$. The $u_{d}= f\left( \Delta T_{cl}, \theta_{micro}\right)$ dependence required for Eq.~\eqref{eq:dot_xcl} is given in \cite{JFM22}. Figure~\ref{fig:OFT_water} demonstrates that the amplitude of oscillations decreases with increasing $\theta_{micro}$, which qualitatively agrees with the experiment. This amplitude reduction can be explained as follows. It is well known \cite{Snoeijer10,JFM22} that the contact line recedes faster for large contact angles. This is evident from Fig.~\ref{fig:OFT_water}, where the contact line dynamics is shown in yellow lines. For large $\theta_{micro}$, the contact line reaches the condenser border very fast and evaporation stops, which causes a smaller oscillation amplitude.

According to Fig.~\ref{fig:OFT_water}, the oscillation frequency decreases with wettability. This is related to the increase of amplitude and thus of average vapor volume $\bar \Omega_v=S\bar x_m$, cf. Eq.~\eqref{omega0}; the increase of $\bar x_m$ is evident in Fig.~\ref{fig:OFT_water}.

\section{Conclusions}

We propose a new one-dimensional physical model to describe the self-sustained oscillations observed in pulsating heat pipes. We call it the Oscillating Film Thickness (OFT) model to distinguish it from previous models. Choosing the one-dimensional approach is a trade-off between the model's accuracy and efficiency, as it can substantially reduce computational time and resources compared to multi-dimensional approaches. The model is based on the solid physical background developed thanks to recent advances in the physical understanding of liquid film dynamics in capillary tubes. Thanks to them, this model has the advantage of being relatively simple and reproducing well the experimental data. The OFT model reflects essential features of liquid film dynamics in real fluid channels much better than previous approaches. Three aspects of liquid film behavior are emphasized: the film deposition by receding liquid menisci, the time variation of film thickness due to phase change, and the contact line receding that shortens the film. The OFT model accounts for the wetting properties and explains the improvement of the PHP performance with the wettability increase, in qualitative agreement with experimental observations.

%outlook
The present work is focused on the PHP hydrodynamic behavior and the validation of the new physical model for liquid films, which is the foundation of accurate PHP simulations in the future. The thermal coupling between the liquid and the solid tube should be accounted for when extending the current model to multi-branch PHP simulations, from which macroscopic quantities of PHP systems will be obtained.

\begin{acknowledgments}
The present work is supported by the project TOPDESS, financed through the Microgravity Application Program by the European Space Agency. This article is also a part of the PhD thesis of X.~Zhang, co-financed by the CNES and the CEA NUMERICS program, which has received funding from the European Union's Horizon 2020 research and innovation programme under grant agreement No 800945 - NUMERICS - H2020-MSCA-COFUND-2017. Additional financial support of CNES awarded through GdR MFA is acknowledged.
\end{acknowledgments}

\appendix
\section{Film shape factor}\label{app:form}

Consider a spatially varying film thickness $\delta(x)$ along the film length $L_{f}$. The rigorous definition of the effective film thickness $\langle \delta\rangle$ comes from the film mass because it enters the above model through the film mass expression \eqref{eq:mf}. Under the assumption $\delta(x)\ll d$, the film mass for the varying $\delta$ is
\begin{equation}\label{mfvar}
m_{f}=\pi \rho_{l} \int_0^{L_{f}} [d-\delta(x)]\delta(x) \text{d}x \approx \pi d \rho_{l} \int_0^{L_{f}} \delta(x) \text{d}x= \pi d \rho_{l} \langle \delta\rangle L_{f},
\end{equation}
and the last expression conforms to Eqs. (\ref{eq:mf}, \ref{Sf}). Therefore, the effective film thickness satisfies
\begin{equation}\label{avd}
\langle \delta\rangle \equiv \frac{1}{L_{f}}\int_0^{L_{f}} \delta(x)\mathrm{d}x.
\end{equation}

As the mass flux is proportional to $\delta^{-1}$, a factor $\varsigma$ needs to be introduced for its calculation as
\begin{equation}\label{eq:gamma}
  \frac{\varsigma}{\langle \delta\rangle} \equiv \frac{1}{L_{f}}\int_0^{L_{f}} \frac{\mathrm{d}x}{\delta(x)}.
\end{equation}

Two simple cases are presented here, to illustrate the $\varsigma$ values for different $\delta(x)$ profiles.

Consider a wedge-like film, i.e. $\delta(x)=ax+b$ in the interval $[0,L_{f}]$; $\delta(0)=\langle \delta\rangle-\Delta\delta$ and $\delta(L_f)=\langle \delta\rangle+\Delta\delta$ are both positive. According to Eq.~\eqref{eq:gamma},
\begin{equation*}
  \varsigma=\frac{\langle \delta\rangle}{2\Delta\delta} \ln\frac{\langle \delta\rangle+\Delta\delta}{\langle \delta\rangle-\Delta\delta}\approx 1+\frac{1}{3} \left(\frac{\Delta\delta}{\langle \delta\rangle}\right)^2,
\end{equation*}
for $\Delta\delta\ll\langle\delta\rangle$. One can see that $\varsigma$ is larger than unity. This is a general tendency as $\varsigma$ presents the ratio of arithmetic and harmonic means which is known to be always larger than unity. This means that the value $\sim 0.5$ used in previous FEC model simulations \cite{IJHMT10} was wrong.

Consider now a more realistic case, where the film is deposited by an oscillatory liquid plug from an immobile contact line. When the plug oscillates with the amplitude $A$, the meniscus speed during the first (receding) half of the period is
\begin{equation*}\label{eq:Um(x)}
 u_{m}(x)=u_{a}\left[\frac{x}{A} \left(2- \frac{x}{A}\right)\right]^{1/2}
\end{equation*}
where $u_{a}$ is the velocity amplitude. Neglecting the mass exchange and assuming that $u_{a}$ is low enough, the film thickness $\delta (x)$ deposited by the meniscus can be approximated by the expression \cite{JFM21},
\begin{equation}\label{eq:delta(x)}
  \delta(x) \approx \delta_a \left[ \frac{x}{A} \left(2- \frac{x}{A}\right)\right] ^{1/3},
\end{equation}
where $\delta_a = 0.67d \left( \mu u_{a}/\sigma \right)^{2/3} $ and $\langle \delta\rangle \approx 0.84 \delta_a$. Substituting Eq.~\eqref{eq:delta(x)} into Eq.~\eqref{eq:gamma} and noting that $L_f=2A$, one obtains numerically $\varsigma \approx 2.18$. One can thus consider that the realistic $\varsigma$ values are between 1 and 3.

\section{Relative contribution of the central meniscus part, the film edge, and the liquid film}\label{app:rel}

\begin{figure}[ht]
  \centering
  \includegraphics[width=6.5cm,clip]{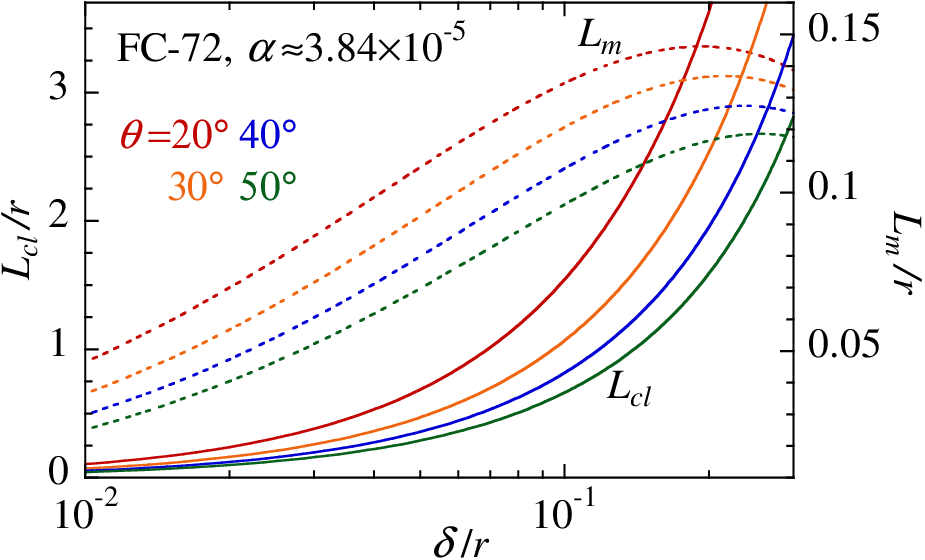}
  \caption{Equivalent film lengths corresponding to the phase change: of  the contact line region $L_{cl}$ (solid lines) and of the central meniscus part $L_{m}$ (dashed lines). The values of $\theta$ are indicated by different colors.}\label{fig:L_d-L_mc}
\end{figure}

Consider the interface geometry of Fig.~\ref{fig:dotm_menisci}a, where a liquid film of effective thickness $\delta$ is attached to the meniscus. The mass exchange from the contact line region, i.e. the film edge, can be estimated with Eq.~\eqref{eq:Jcl(phi)} as $J_{cl}=J(\phi)$, where $\phi$ is determined from $\delta$ with Eq.~\eqref{delta}.  By using Eq.~\eqref{eq:Jf}, one can calculate a film length
$L_{cl}$ that would produce the evaporation rate $J_f$ equal to the evaporation rate $J_{cl}$ from the contact line region under the same superheating. The result is $$L_{cl}=\frac{rw(\delta)\delta}{\varsigma(r-\delta)},$$ where it is more convenient to reason in terms of $\delta$ than $\phi$.
Similarly, one can estimate the mass exchange $J_{m}=J(\pi/2)-J(\phi)$ from the meniscus central part and an equivalent film length $L_{m}$:
$$L_{m}=\frac{r[W-w(\delta)]\delta}{\varsigma(r-\delta)}$$
Fig.~\ref{fig:L_d-L_mc} plots both $L_{cl}$ (the left $y$-axis) and $L_{m}$ (the right $y$-axis) for FC72, $r=\SI{1}{mm}$ and $\delta$ ranging from 10
to \SI{200}{\micro \meter}. For $\theta=20^\circ$ and $\delta=\SI{50}{\micro m}$, we obtain  $L_{cl}\approx\SI{0.70}{mm}$ and $L_{m}\approx\SI{0.11}{mm}$. Both these values are much shorter than typical film length, of the order of centimeters. One concludes that while the contribution to overall mass flux from the contact line regions is several times larger than the contribution from central part of menisci, both are usually much smaller than the mass flux from flat films (if a film is present).

\section{Singularity appearing when a meniscus appro\-aches the contact line}\label{appAsympApp}

Consider a possible divergence of the effective film thickness $\delta$ when the meniscus approaches the contact line. In this case, $u_{m}-u_{cl}<0$, $u_{m}<0$, but $u_{cl}>0$. The film collection rate is given by Eq.~\eqref{eq:dot_mdep2}, i.e. $\dot m_{dep} =S_{f}u_{m} \rho_{l}$. Since the film thickness is large, it is possible to neglect both $J_{f}$ and $J_{cl}$ terms with respect to the term $\dot m_{dep}$ in Eq.~\eqref{eq:dot_mf}. By assuming $\delta\ll d$, one obtains from Eqs.~(\ref{eq:mf}-\ref{Sf})
\begin{equation}\label{eq:c2_1}
\dot \delta=\frac{\delta u_{cl} }{x_{m} -  x_{cl}}>0.
\end{equation}

Close to the instant $t_0$ of the film disappearance $x_{m}\to x_{cl}$, one can assume a constant relative meniscus velocity; for $t\lesssim t_0$, $x_{m} = x_{cl}+(u_{m} - u_{cl})(t-t_0)$. Eq.~\eqref{eq:c2_1} becomes
\begin{equation}\label{eq:c2_2}
 \dot \delta(t)=\alpha\frac{\delta}{t_0-t}
\end{equation}
with
$$\alpha=-\frac{u_{cl} }{u_{m} - u_{cl}}>0.$$
This results in $\delta(t)\sim(t_0-t)^{-\alpha}$, so $\delta$ diverges. However, since, typically, $u_{m} \gg u_{cl}$, the divergence remains weak.

\bibliography{PHP,Taylor_bubbles,ContactTransf}

\end{document}